\documentclass[12pt]{article}
\DeclareUnicodeCharacter{0301}{\'{e}}
\usepackage{amssymb}
\usepackage{amsmath}
\usepackage{amsthm}
\usepackage{color}
\usepackage{comment}
\usepackage{enumitem}		
\usepackage{geometry}		
\usepackage{graphicx}
\usepackage{authblk}
\usepackage{amssymb}
\usepackage{listings}
\usepackage{caption}
\usepackage{algorithm}
\usepackage{algpseudocode}

\usepackage{subcaption}
\usepackage{adjustbox}
\usepackage[table,xcdraw]{xcolor} 
\usepackage{array} 
\usepackage{makecell} 

\makeatletter
	\@addtoreset{equation}{section}
\makeatother

\let\originalleft\left
\let\originalright\right
\renewcommand{\left}{\mathopen{}\mathclose\bgroup\originalleft}
\renewcommand{\right}{\aftergroup\egroup\originalright}

\newlist{romanlist}{enumerate}{3}
\setlist[romanlist]{label=\roman*),ref=(\roman*)}

\begin{document}

\newcommand{\cF}{\mathcal{F}}
\newcommand{\cP}{\mathcal{P}}
\newcommand{\cR}{\mathcal{R}}
\newcommand{\cS}{\mathcal{S}}
\newcommand{\cT}{\mathcal{T}}
\newcommand{\ee}{\varepsilon}
\newcommand{\rD}{{\rm D}}
\newcommand{\re}{{\rm e}}

\newtheorem{theorem}{Theorem}[section]
\newtheorem{corollary}[theorem]{Corollary}
\newtheorem{lemma}[theorem]{Lemma}
\newtheorem{proposition}[theorem]{Proposition}

\theoremstyle{definition}
\newtheorem{definition}{Definition}[section]


\title{Path planning for multi-quadrotor 3D boundary surveillance using non-autonomous discrete memristor hyperchaotic system}

\author[1]{Harisankar R} 
\affil[1]{School of Electronic System and Automation,\\ Digital University Kerala\\
Thiruvananthapuram, PIN 695317, Kerala, India \newline
E-mail: harisankarrj@gmail.com}

\author[2]{Abhishek Kaushik}
\affil[2]{School of Computer Science and Engineering,\\ Digital University Kerala\\
Thiruvananthapuram, PIN 695317, Kerala, India \newline
E-mail: 	kaushik.or.abhishek@gmail.com}
            
\author[3]{Sishu Shankar Muni}
\affil[3]{School of Digital Sciences,\\ Digital University Kerala\\
Thiruvananthapuram, PIN 695317, Kerala, India \newline
E-mail: sishushankarmuni@gmail.com}
\maketitle

\begin{abstract}
Recent studies have shown that chaotic maps are well-suited for applications requiring unpredictable behaviour, making them a valuable tool for enhancing unpredictability and complexity. A method is developed using 3D parametric equations to make boundary surveillance more robust and flexible to effectively cover and adapt to the ever-changing situations of boundary surveillance. A non-autonomous discrete memristor hyperchaotic map is utilized to significantly enhance the unpredictability of trajectories in boundary surveillance.
Since the scaling property is an affine transformation, and the hyperchaotic systems are invariant under affine transformations, the adaptability of hyperchaotic trajectories is visually demonstrated by applying the scaling property for robust path planning. Furthermore, hyperchaotic systems enable the generation of mutually independent trajectories with slight variations in initial conditions, allowing multiple quadrotors to operate simultaneously along a single guiding path. This minimizes the chance of overlap and interference, ensuring effective and coordinated surveillance.
\end{abstract}
\section{Introduction}
\label{Introduction}
Recently, quadrotors have become popular for surveillance missions due to their unmatched manoeuvrability, ease of deployment, and reduced mechanical complexity compared to other aerial vehicles \cite{32}. Their compactness and agility make them very apt to fly in complex environments, thus suitable for open and confined spaces. Equipped with advanced sensors and cameras, quadrotors can capture high-resolution images and videos, thereby providing an all-detailed and comprehensive coverage of the area under surveillance. Compared with conventional helicopters, quadrotors result in a rise in reliability and a significant reduction in the costs for manufacturing, operation and maintenance. Such characteristics outlined quadrotors more promising in a wide range of unmanned military and civilian missions, like border patrol\cite{29}, search and rescue \cite{30}, and infrastructure monitoring\cite{31}. Similarly, in earlier works, researchers have designed various surveillance robots, leveraging advanced technology for inspection and monitoring using ground robots \cite{1}, intelligent obstacle avoidance through neural networks \cite{2}, and real-time monitoring systems in environmental and industrial applications \cite{3}. Other applications of such robots encompass underwater surveillance \cite{4}, smart homes \cite{5}, power grid line inspection \cite{6} and military surveillance utilizing unmanned aerial vehicles (UAVs) \cite{7}.

The significant advancements taking place in unmanned aerial vehicles are reviewed in \cite{8}, considering developments in architectural design, control algorithms, and technologies for performance enhancement and stability. For enhancing manual quadrotor operation, solving some of the critical challenges like precise trajectory tracking \cite{9}, vibration reduction \cite{35}, and robustness against disturbances \cite{10} are critical. Initially, the quadrotor creates thrust from the changing speeds of its four rotors, allowing it to control its movement and maneuver quite precisely in three dimensions. Later, equipped with sophisticated algorithms and sensors, navigation, obstacle avoidance, and real-time decision-making enabled a UAV to operate without human intervention. 
Recent development in autonomous quadrotors has been achieved by the development of a kinodynamic path search combined with B-spline optimization, such that smooth, feasible, and time-efficient trajectories are able to be generated \cite{11}. It drives quadrotors to fly in complex three-dimensional scenes with better precision and reliability, hence helping to increase their capability to decide in real time and avoid obstacles in dynamic conditions.  Later research led to the addition of other purposeful features in surveillance, specifically navigation \cite{12,13,14} and localization even when in a swarm \cite{15}, object or location tracking 
\cite{16}, multiagent coordination of UAVs \cite{34} and more. Quadrotors utilize a monocular camera and minimal sensors to improve ground target tracking in outdoor environments, aiming to decrease system uncertainties and disturbances at the same time \cite{16}.

Surveillance missions can, in general, be divided into three categories: point, regional and boundary surveillance \cite{17}. Earlier boundary surveillance, mostly done by humans, can be effectively handled by the discussed robots above. Missions of surveillance can be categorized into point surveillance - which regards a small and localized area or an object, and regional surveillance, which concerns larger areas. And boundary surveillance dealing with borders or perimeters. 

This paper focuses on the quadrotor-based boundary surveillance problem. Traditional surveillance methods generally rely on somewhat predictable patterns and are open to detection and countermeasures. So the relevance of understanding and managing the unpredictability in flight path motion has become prominent. The unpredictability makes it hard for adversaries to foretell and avoid detection and ensures better coverage of the surveillance area. It can be realized with
divergence of the sequential points in chaotic system \cite{18}.

There is a four-scroll chaotic attractor by Akgul et al. \cite{19} described by three coupled nonlinear differential equations, where chaotic behaviour is observed by setting specific parameters to zero. And analyzed these chaotic dynamics by using Lyapunov exponents, adaptive control, and circuit simulations and further validated their findings with experimental data in order to explore applications in various scientific and engineering fields. Chaotic signals could also be applied in image processing for secure image encryption to add security against unauthorized access by persons\cite{61}. Later Akgul et al.\cite{20} provided contributions for 3D chaotic systems without equilibrium points to improve encryption techniques and secure communications, providing video encryption solutions that fulfil international security requirements. Zang et al.\cite{21} given an overview of the chaos and fractal applications to robotics, showing their role in the improvement of mobile robot behaviour, optimization algorithms, bipedal locomotion, and modular robotic mechanisms.

In a recent article, Nwachioma et al.\cite{22} presented a new chaotic system with four attractors, including fixed point and symmetric chaotic strange attractors, which are used as control input of a differential drive mobile robot to enable unpredictability and complete scanning of the workspace. The space-filling property of the chaotic system is used to create paths that systematically cover the entire workspace and are utilized quite well by the above robot \cite{22}. This is particularly valuable for tasks requiring thorough area coverage, such as search and rescue, surveillance, and autonomous exploration. Nasr et al.\cite{23} contributed a multi-scroll chaotic system which improves efficiency in trajectory planning, maintains operational boundaries, integrates control with planning, and works with larger workspaces. Then a multi-direction multi-scroll chaotic system model\cite{24} is presented along with its dynamics, FPGA implementation, and cryptographic application. In addition, experimental results have shown that the proposed system \cite{24} can generate highly sensitive chaotic sequences that are very effective for generating highly sensitive chaotic sequences in the case of robust image encryption.

A chaotic system may be understood as one whose behaviour is highly divergent and, at the same time, seemingly random, even though it is deterministic. All this is due to sensitivity to initial conditions, which ensure high security, making encrypted data very sensitive to the key and plaintext\cite{24}. Additionally, the unpredictable behaviour of chaotic systems is utilised in trajectory planning for mobile robots to achieve complete and non-repetitive area coverage\cite{22,23}.
Another relevance of chaotic trajectories in surveillance is that they evade forming any predictable pattern, making the mission of surveillance both stealthy and efficient.

In our application, we rely heavily on the property of unpredictability. While generating random points also gives unpredictability, there is a possibility that the same points or areas might get revisited repeatedly and also have the chance to happen consecutively \cite{62}. So, it is not efficient in applications that require unpredictable trajectories. Chaotic trajectories provide the required unpredictability while avoiding these consecutive overlaps. Also, apart from random prediction, chaotic attractors always fall within a chaotic pattern based on the system chosen. Further, chaotic trajectories are highly sensitive to initial conditions and system parameters, meaning minute changes in these initial conditions of the system can lead to entirely different dynamical behaviours. This parametric dependence is important because it provides an option to fine-tune the trajectories to satisfy particular surveillance necessities, such as ensuring thorough coverage of critical areas or avoiding certain regions. Even a minute change in initial conditions is enough if the target is to get a different trajectory that is mutually independent from the previous one. The capability to adjust parameters in real-time gives adaptability and flexibility to chaotic systems, which enables drones to respond effectively to changing conditions like moving obstacles or varying surveillance priorities.

The algorithm proposed in \cite{22,23} offers a solution for path planning of robots in 2D applications only. However, this work addresses the boundary surveillance problem in three-dimensional space. Also, this paper is an upgrade of the earlier work done by P.S. Gohari et al.\cite{26}, who utilized chaotic trajectories to enhance quadrotor performance in similar surveillance tasks. Some of the limitations of the above papers are: the definition of the closed contour was in a 2D plane, thus it constrained the quadroter vicinity within that closed contour. Since there is one technique that involves taking predefined points at the edges of real-world boundaries to form a definite contour. The method described in \cite{26} does not offer a mechanism to define the closed contour for surveillance. 

The approach in\cite{26} does not account for variations in elevation, and the absence of a vertical component within the trajectory reduces the system's adaptability to diverse environments. A 2D contour does not allow effective navigation around three-dimensional structures such as buildings, towers, variations in terrain, or complex environments like dense urban areas, industrial facilities, and densely forested regions. These scenarios often involve complex and dynamic features, requiring advanced algorithms capable of real-time adaptation and vertical movement for precise path planning and obstacle avoidance. From this point of view, the quadrotor can miss important areas, perspectives and other details. so it may not deliver an effective surveillance application. This paper solves the above problem and improves quadrotor surveillance over the boundary in three-dimensional space by replacing the  2D closed contour\cite{26} with a vertically extended open helical contour.

Another significant limitation of the paper \cite{26} is using the Henon map as a chaotic attractor to derive chaotic trajectories to navigate quadrotors. The problem associated with the Henon equations is that it creates a chaotic attractor with limited occupation in the trapping region, resulting in a chaotic map with less spatial coverage \cite{26}. To fill this gap, a hyperchaotic map was used to address this and improve the unpredictability while substantially increasing the spatial coverage compared to the Henon map.
Hyperchaotic attractors are more complex in dynamics when compared to chaotic ones because the former has more than one positive Lyapunov exponent while the latter has only one positive Lyapunov exponent\cite{65}. So it is able to significantly support the generation of mutually independent trajectories by using minute variations in the initial conditions of hyperchaotic systems, which ensures that more quadrotors can fly simultaneously on a single guiding contour. Here are some applications showcasing the advantages of hyperchaotic systems for better security and unpredictability in biometric authentication\cite{27} and image encryption\cite{28}, leveraging their complexity in dynamics to improve resistance against various forms of attacks. Gohari et al.\cite{26} mentioned that the chosen henon system was simple and efficient due to the limitations of quadrotor processors and power resources. However, our approach uses a hardware realizable hyperchaotic system that offers enhanced performance and efficiency for real-time applications than any other options. Further details are provided in the subsequent sections.
\\

The main contributions are:
\begin{itemize}
    
    \item The upgradation of the traditional 2D contour to a helical 3D contour in path planning greatly improves the effectiveness and unpredictability of the quadrotor's boundary surveillance, guaranteeing greater coverage, enhanced adaptability to navigate around vertical structures like buildings, and optimized monitoring capabilities in three-dimensional environments with a wider range of visual information.
    \item The usage of a hyperchaotic system instead of other chaotic systems significantly improves spatial coverage and the degree of unpredictability in the trajectory generated. Thus, it ensures complex exploration and rendering effectiveness in real-time applications.
    \item The hardware realizable hyperchaotic system enhances its practical application in real-time quadrotor surveillance with cutting-edge performance and efficiency.
    
    \item The use of hyperchaotic systems allows for the generation of mutually independent chaotic trajectories. The minute variations in initial conditions enable multiple quadrotors to operate simultaneously on a single contour,  significantly reducing the likelihood of overlap and interference between quadrotors operating on the same contour.
    
\end{itemize}

The remaining part of the paper is organized as follows: section 2 discusses the relevance of the hyperchaotic non-autonomous discrete memristor attractor, providing a parametric comparison with other chaotic systems. Section 3 discusses the methodology of generating 3D trajectories using hyperchaotic points. In section 4, the quadrotor model is used to trace the desired trajectories.  Section 5 shows the results and simulations, showcasing the graphical validation of generating multiple hyperchaotic trajectories that enable multiple quadrotors to simultaneously trace these paths. Section 6 outlines the practical implications of deploying quadrotors, which follow the algorithm discussed in the methodology. Final section 7 summarizes the major findings of the paper and gives avenues for further research.

\begin{figure}[!ht]
    \centering
    \begin{subfigure}[t]{0.49\textwidth}
        \centering
        \includegraphics[width=\textwidth]{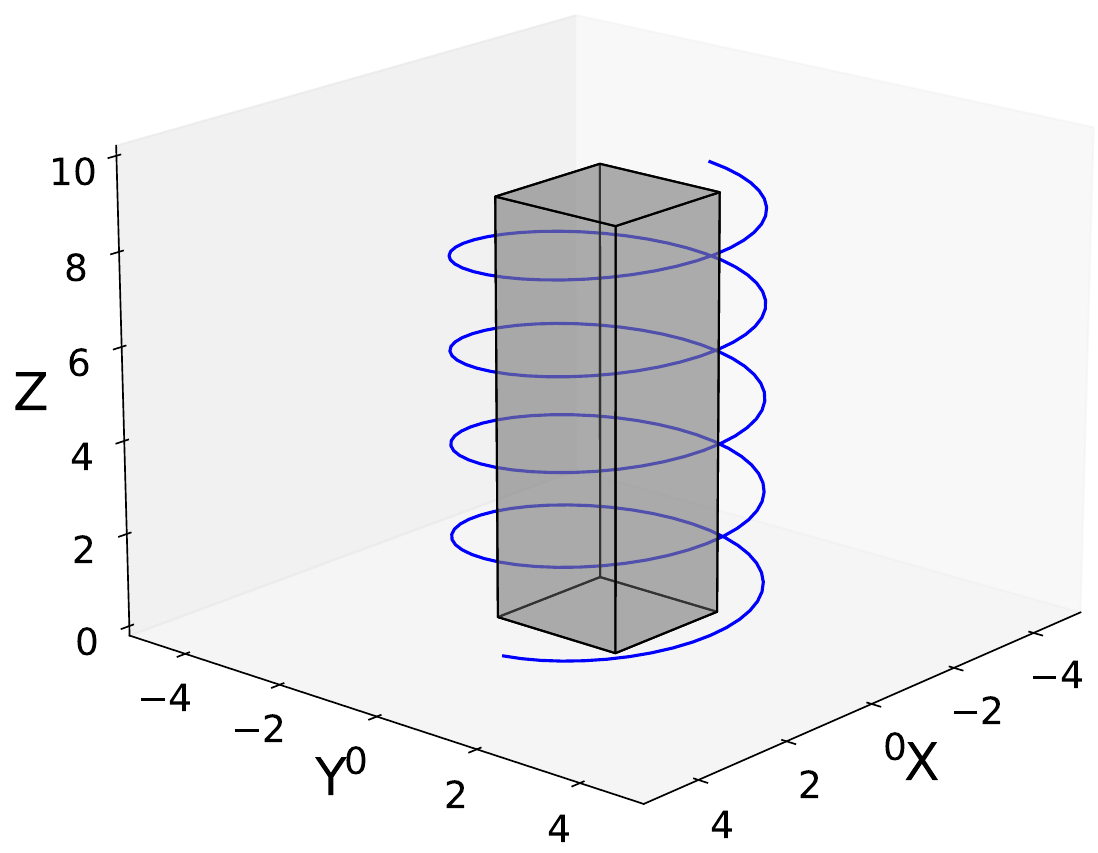}  
        \caption{}
        \label{fig:a}
    \end{subfigure}
    \hfill
    \begin{subfigure}[t]{0.5\textwidth}
        \centering
        \includegraphics[width=\textwidth]{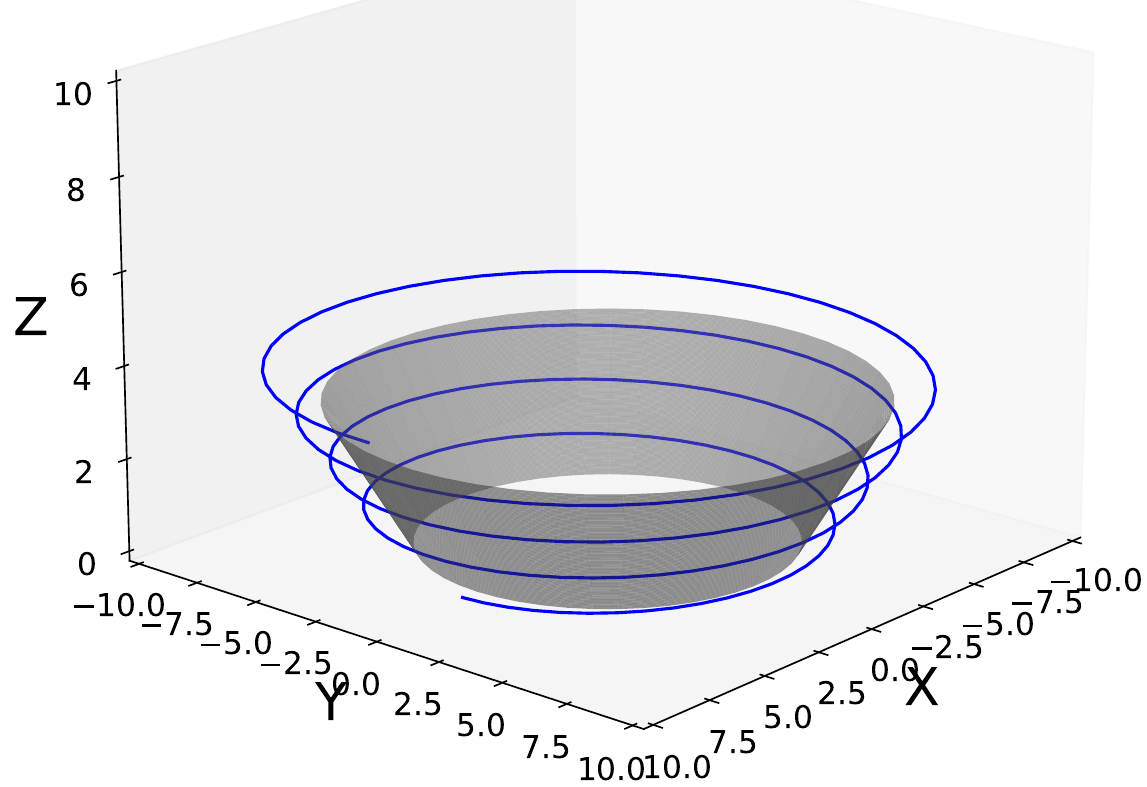}  
        \caption{}
        \label{fig:b}
    \end{subfigure}
    
    \vskip\baselineskip
    \begin{subfigure}[t]{0.55\textwidth}
        \centering
        \includegraphics[width=\textwidth]{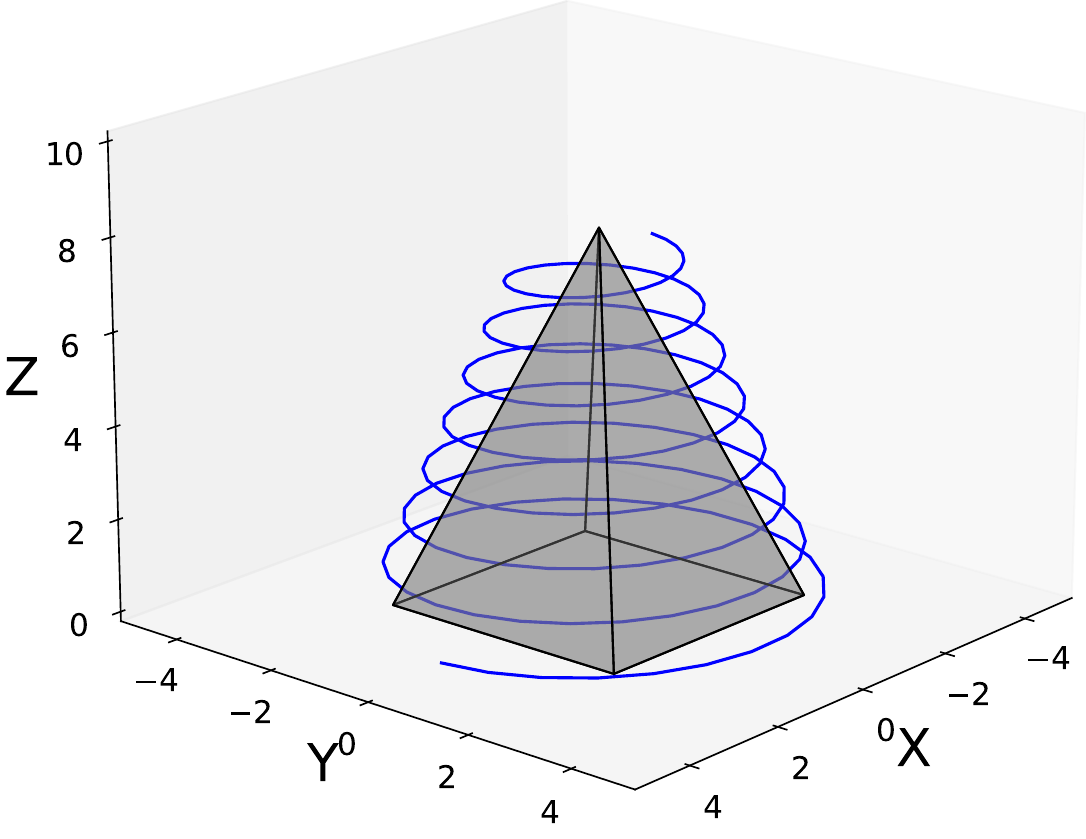}  
        \caption{}
        \label{fig:c}
    \end{subfigure}
    
    \caption{Illustration of various open helical contours in three-dimensional space, which can be utilized for three structural navigation scenarios. a) cylindrical helical contour whose radius is constant. b) Helical contour whose radius is incrementing. c) Helical contour whose radius is decrementing. }
    \label{fig:1}
\end{figure}

\section{Description of the problem}
\label{Description of the problem}

Chaotic maps play a key role in surveillance systems due to their properties, which give invaluable advantages against traditional methods of random point generation, as discussed earlier. Apart from the closed contour confined in the 2D plane \cite{26}, this paper introduces a helical contour that progresses around vertical structures or buildings based on a pitch value. The pitch refers to the distance in a vertical direction between successive turns of the helix and hence determines how tightly or loosely the contour wraps around the structure as it ascends.

Figure.\ref{fig:1} illustrates the concept of the helical contour progressing vertically upwards along the z-axis.
Fig.\ref{fig:a}.  depicts a general or cylindrical helix[36] whose radius is constant, or the tangent of the helix makes a constant angle with a fixed line in space, which makes it suitable for uniform ascension. And this open helical contour is swirling around a vertical building. Additionally, Fig.\ref{fig:b} contains a helix with an incrementing radius, which is particularly useful for navigating wide or expansive structures like stadiums. Conversely, Fig.\ref{fig:c} is a helix with radius decrementing, ideal for structures like pyramids, where the contour naturally tightens as it ascends. The blue line represents an open helical contour in three-dimensional space, which is designated to be survialenced by the quadrotor. This contour serves as a boundary for the quadrotor to follow during ascending or descending around these structures to ensure full coverage during surveillance.

In the domain of path planning for mobile robots, chaotic systems are mainly utilized in two ways, which are the generation of unpredictable trajectories and workspace coverage using chaotic-based optimization algorithms. This paper relies on the first method, which uses the inherent unpredictability of the system to generate trajectories that are quite hard to predict. In the surveillance problem discussed, a chaotic system generating unpredictable trajectories from two different nearby initial conditions would guarantee mutually independent paths. The unpredictability of trajectory ensures that the surveillance area is covered without repetition and with comprehensiveness. The coverage of chaotic paths over the entire surveillance region without leaving a significant gap is referred to as space-filling property. Nwachioma et al.\cite{22} utilized this property to enable mobile robot for thorough scanning of the workspace.

To enhance the degree of unpredictability in generated trajectories, we adopted a hyperchaotic system instead of a chaotic system as in \cite{26}. Along with enhancing unpredictability, the objective is to show how sensitivity to the initial conditions provides mutually independent trajectories for multiple quadrotors. 

\subsection{2D non-autonomous discrete memristor-based hyperchaotic map}

\begin{figure}[ht]
    \centering
    \includegraphics[width=0.6\textwidth]{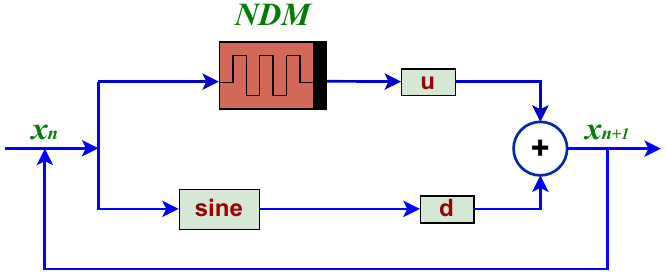}
    \caption{Schematic of framework to produce NDMH map, which combines NDM, control parameters and sine map. }
    \label{fig:sch_NDM_map}
\end{figure}

This paper introduced hyperchaotic systems to increase unpredictability, which would in turn, yield multiple trajectories. To our knowledge, implementing hyperchaotic systems for multiple trajectories in real-world applications remains uncharted territory.

To achieve a higher degree of unpredictability and rate of divergence of the nearby sequence of points, we have used the hyperchaotic attractor from a non-autonomous discrete memristor (NDMH) map. In general, the existence of hyperchaos requires at least four dimensions in continuous dynamical systems\cite{40,41}, while for discrete dynamical systems, it is possible to obtain it with two dimensions\cite{42,43}.In particular, continuous memristor-based systems in the $(\psi, q)$  domain cannot reach hyperchaos because they only have three dimensions\cite{44}. but they can be hyperchaotic in the $(v, i)$ domain with four dimensions. However, 2D memristor-based discrete maps are capable of generating hyperchaos\cite{43,45,46}. This makes discrete systems advantageous over continuous ones in terms of less complex algebraic equations \cite{47}and higher computational efficiency, which can be used for chaos-based industrial applications\cite{48}.

From the circuit theory perspective, a memristor is a two-terminal nonlinear device that shows a pinched hysteretic loop in the voltage-current plane under periodic input by voltage $v(t)$ or current $i(t)$. An ideal charge-controlled memristor \cite{49} is given by

\begin{equation}
\left\{
\begin{aligned}
v(t) = M(q) \cdot i(t)\\
\frac{dq(t)}{dt} = i(t)
\end{aligned}
\right.
\label{eq:ideal_memristor}
\end{equation}

\begin{figure*}[ht!]
    \centering
    \begin{subfigure}[t]{0.45\textwidth}
        \centering
        \includegraphics[width=\textwidth]{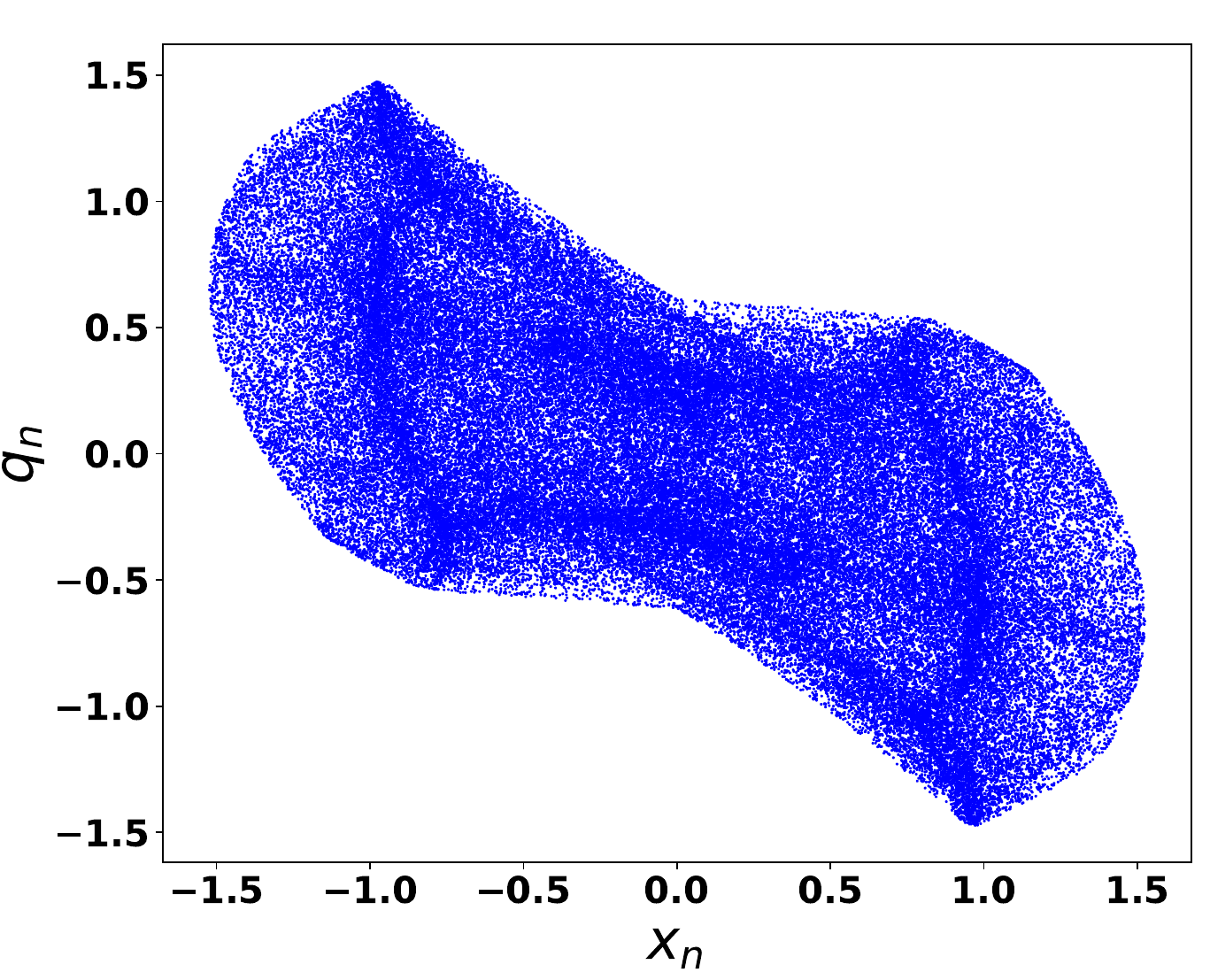}  
        \caption{}
        \label{fig:NDMH_map}
    \end{subfigure}
    \hspace{10pt} 
    \begin{subfigure}[t]{0.45\textwidth}
        \centering
        \includegraphics[width=\textwidth]{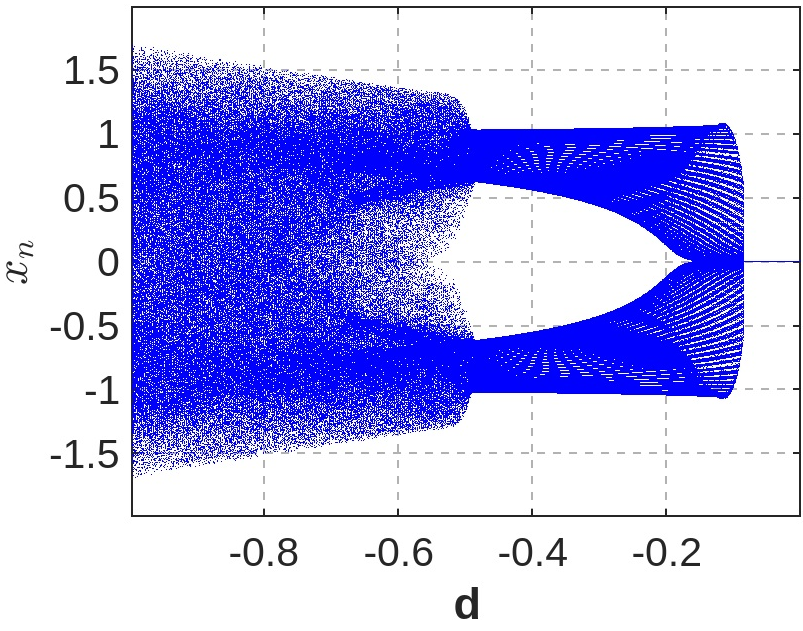}  
        \caption{}
        \label{fig:Bif_pltvsd}
    \end{subfigure}
    
    \vskip\baselineskip
    \begin{subfigure}[t]{0.49\textwidth}
        \centering
        \includegraphics[width=\textwidth]{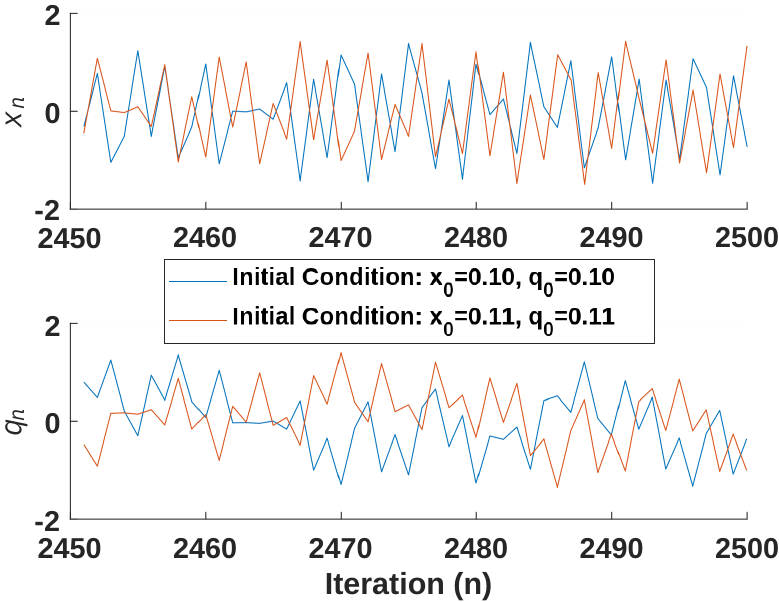}  
        \caption{}
        \label{fig:xqvsn}
    \end{subfigure}
    \hspace{5pt} 
    \begin{subfigure}[t]{0.45\textwidth}
        \centering
        \includegraphics[width=\textwidth]{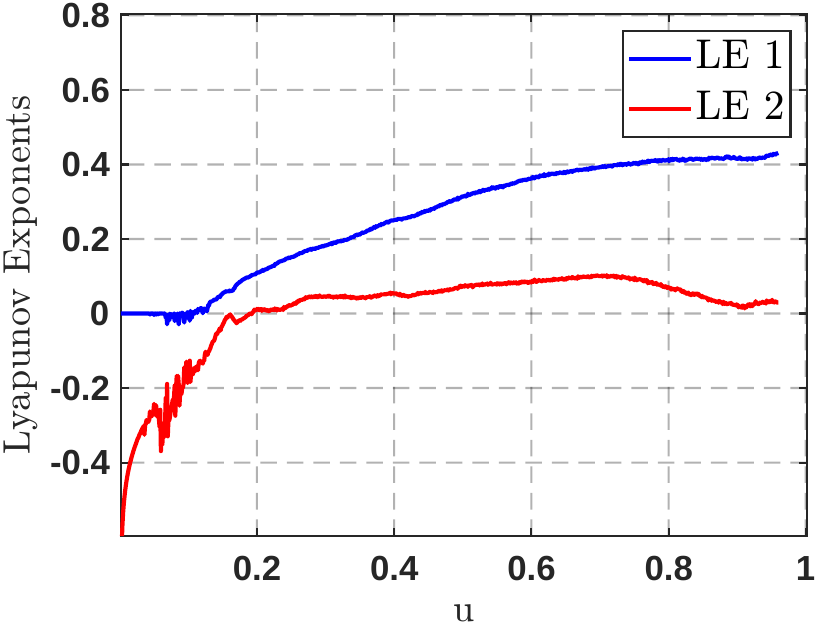}  
        \caption{}
        \label{fig:LE}
    \end{subfigure}
    
    \caption{Dynamics of the NDMH Map - (a) Hyperchaotic attractor for specific parameters and initial conditions. (b) Bifurcation diagram of \(x_n\) versus \(d\). (c) Sensitivity to initial conditions, with \(x_n\) and \(q_n\) plotted against iterations \(n\). (d) Lyapunov exponents versus \(u\), confirming hyperchaotic behavior.}
    \label{fig:main_figure}
\end{figure*}

where q is used to denote the variable for charge, and $M(q)$ denotes
the memristance in Ohms.\\
The continuous memristor model can be discretized using the forward Euler difference method\cite{45,46}. Let ${(v_n, i_n, q_n)}$ be the sampled values of voltage $v(t)$, current $i(t)$, and charge $q(t)$ at the $n$-th iteration and $q_{n+1}$ is the sampling value at the ${(n+1)}$-th iteration. Bao et al.\cite{50} modelled a charge-controlled discrete memristor (DM), and that can be expressed as

\begin{equation}
\left\{
\begin{aligned}
v_n = M(q_n) \cdot i_n\\
q_{n+1} = q_n + k.i_n
\end{aligned}
\right.
\label{eq:DM_eqn}
\end{equation}
where $M(q_n)$ represents the sampling value of memristance $M(q)$ at the $n^{th}$ iteration and k=1 .

Inspired by Deng et al.\cite{51}, a new memristance function: {\( M(q_n) = q_n^2 - 1 + c \cdot \sin(r \cdot n + r) \)}
in the ideal discrete memristor Eq.~\ref{eq:DM_eqn}. Consequently, the new NDM mathematical model can be expressed as follows

\begin{equation}
\begin{cases}
v_n = \left[ q_n^2 - 1 + c \cdot \sin(r \cdot n + r) \right] \cdot i_n \\
q_{n+1} = q_n + k \cdot i_n.
\end{cases}
\label{eq:NDM_eqn}
\end{equation}

Then, a new (non-autonomous discrete memristor hyperchaotic) NDMH 2D map is built using the NDM model above Eq.\eqref{eq:NDM_eqn} and the sine map. A schematic diagram for the framework of the NDMH map is shown in Fig.\ref{fig:sch_NDM_map}. The mathematical model of a new 2D NDMH map is:

\begin{equation}
\begin{aligned}
\begin{cases}

x_{n+1} &= u \cdot x_n \cdot \left[ q_n^2 - 1 + c \cdot \sin(r \cdot n + r) \right] \\
&\quad + d \cdot \sin(\pi \cdot x_n) \\
q_{n+1} &= q_n + k \cdot x_n
\end{cases}
\end{aligned}
\label{eq:NDM_map}
\end{equation}

where \( c \), \( k \), \( u \), \( r \), and \( d \) are control parameters. When the parameters are set to \( c = 0.56 \), \( u = 0.8 \), \( k = 1 \), \( r = 0.05 \), \( d = -0.8 \), and the initial value is \( (x_1, q_1) = (0.1, 0.1) \) as per \cite{52}, the hyperchaotic attractor of the NDMH map is shown in Fig.\ref{fig:NDMH_map}.Fig.\ref{fig:Bif_pltvsd} shows the bifurcation plot of $x_n$ versus d and shows higher hyperchaotic behaviour for -0.5  $\leq d \leq -1$, while other parameters and initial conditions remain unchanged. Fig.\ref{fig:xqvsn}  presents the plots of $x_n$ and $q_n$ versus the iteration number n, illustrating its sensitivity to initial conditions, in which the red and blue lines correspond to two initial conditions that are very close to each other. To see the divergence more clearly, the transients were discarded, and the extent of n ranging from 1950 to 2000  was selected. While hyperchaotic systems are known to have two positive Lyapunov exponents, Fig.\ref{fig:LE} shows the Lyapunov exponents (LEs) versus
u, ranging from 0 to 1. At \( u = 0.8 \), the corresponding LEs for  2D NDMH map are \( \text{LE}_1 = 0.411 \) and \( \text{LE}_2 = 0.062 \).In order to make this concrete, the following Table.\ref{table:chaotic_maps} presents performance comparisons for chaotic sequences of some 2D maps. It is clear that the NDMH map is considerably outperforming the others, yielding much more divergent trajectories, which are crucial for our application. And the taken
existing chaotic maps are the Hénon map \cite{53}, Lozi map \cite{54}, hidden NF$_{1\alpha}$  map \cite{56},
NEM$_{1}$  quadratic map \cite{57}, and Sine boostable map \cite{58}.

\begin{table}[htbp]
\centering
\caption{Comparison of Lyapunov exponents (LE\textsubscript{1}, LE\textsubscript{2}) for different chaotic maps with their respective parameters and initial conditions.}
\label{table:chaotic_maps}
\textbf{Table 1}
\fontsize{6}{12}\selectfont 
\setlength{\tabcolsep}{3pt} 
\renewcommand{\arraystretch}{1.2} 

\begin{tabular}{|l|l|l|c|}
\hline
\rowcolor{blue!20}
\textbf{Chaotic Maps} & \textbf{Parameters} & \textbf{Initials} & \textbf{LE$_1$, LE$_2$} \\ \hline
\rowcolor{yellow!20}

Q-DM map & 1.78 & ($-0.5$, $0.5$) & $0.2692$, $0.0925$ \\ \hline
\rowcolor{white}
Hénon map & $(1.4, 0.3)$ & $(0, 0)$ & $0.4208$, $-1.6248$ \\ \hline
\rowcolor{yellow!20}
Lozi map & $(1.7, 0.5)$ & $(0, 0)$ & $0.4701$, $-1.1632$ \\ \hline
\rowcolor{white}
CF$_{\alpha}$ map & $(1.2, 2)$ & $(0.27, 0.28)$ & $0.1344$, $-0.1383$ \\ \hline
\rowcolor{yellow!20}
NF$_{1\alpha}$ map & –- & $(-0.93, -0.44)$ & $0.0599$, $-0.3223$ \\ \hline
\rowcolor{white}
NEM$_{1}$ map & 2 & $(1.7, -0.39)$ & $0.1167$, $-0.2679$ \\ \hline
\rowcolor{yellow!20}
Sine map & $(1.5, 3.8)$ & $(-2, 1)$ & $0.5316$, $-0.7267$ \\ \hline
\rowcolor{white}
NDMH map & \makecell{c=0.56, u=0.8\\k=1, r=0.05\\d=-0.8} & \makecell{0.1, 0.1} & $0.4112$, $0.0625$ \\ \hline
\end{tabular}

\end{table}

\subsection{Kinematic relative motion}

\begin{figure}[ht!]
    \centering
    \begin{subfigure}[t]{0.45\textwidth}
        \centering
        \includegraphics[width=\textwidth]{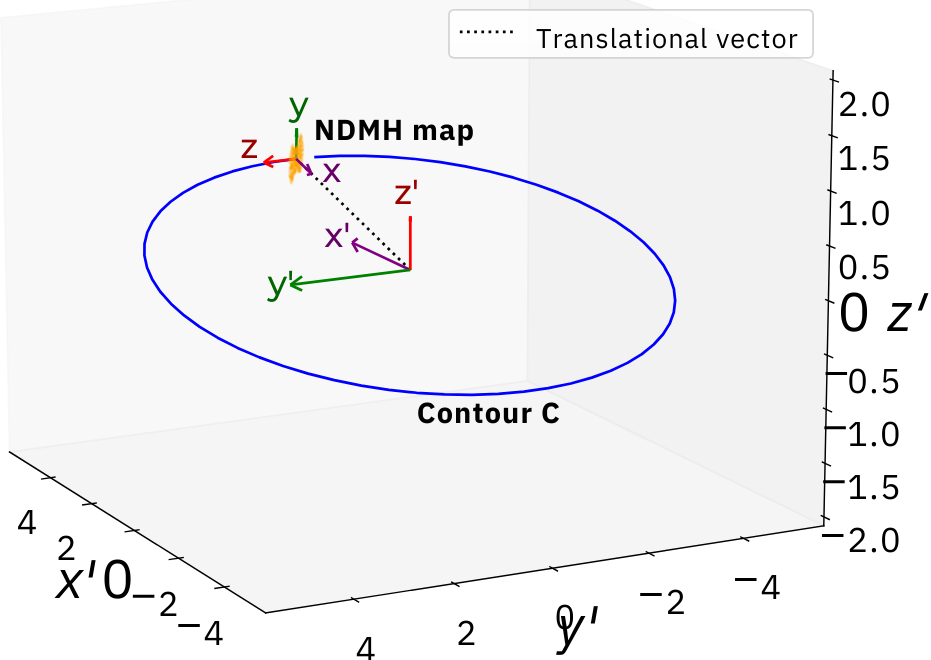}  
        \caption{}
        \label{fig:4a}
    \end{subfigure}
    \hfill
    \begin{subfigure}[t]{0.45\textwidth}
        \centering
        \includegraphics[width=\textwidth]{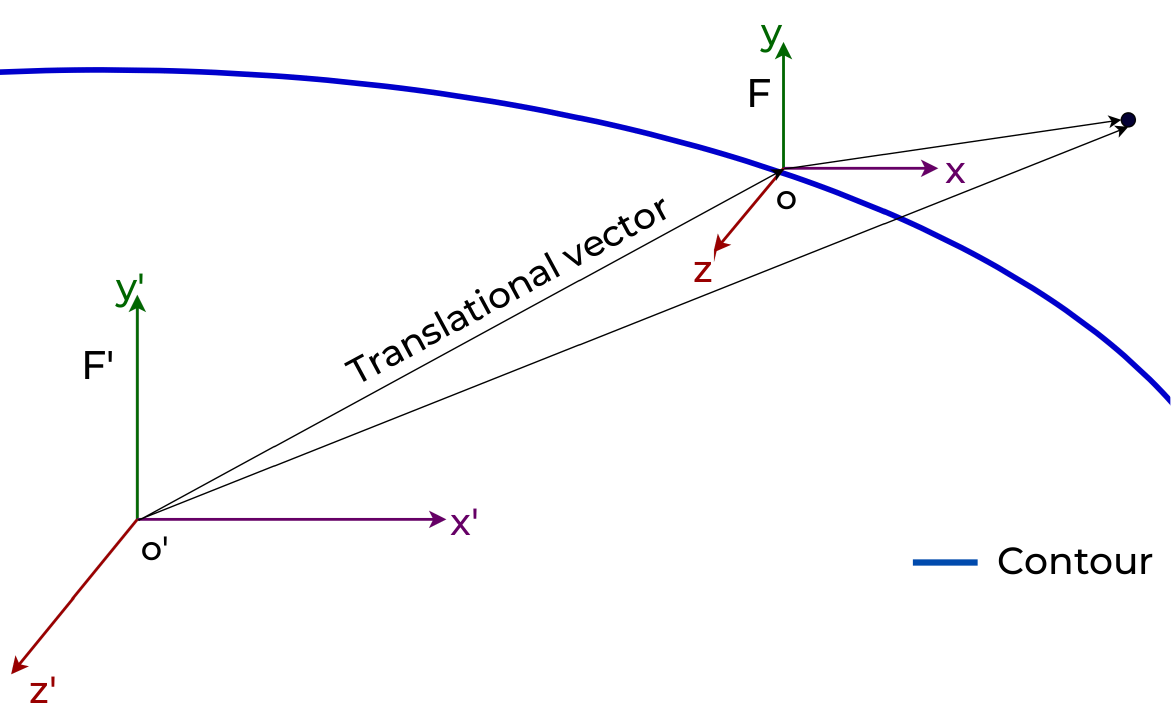}  
        \caption{}
        \label{fig:4b}
    \end{subfigure}

     \caption{Affine transformation from fixed frame $\mathcal{F'}$ to moving frame $\mathcal{F}$. (a) NDMH 2D hyperchaotic map plotted in the xy plane of the moving frame with a 3D elliptical contour. (b) Translational vector \( t_n \) representing the origin of the moving frame $\mathcal{F}$ with respect to the fixed frame $\mathcal{F'}$.}
    \label{fig:4}
\end{figure}
In this paper, analysis is done in a moving frame where the quadrotor is considered stationary. To generate hyperchaotic trajectories for surveillance purposes, the 2D hyperchaotic map needs to be plotted with respect to the moving frame, ensuring that the hyperchaotic behaviour is accurately represented in that frame. Here in Fig.\ref{fig:4a}, the NDMH 2D map is plotted on the xy plane of the moving frame, where the quadrotor is stationary. The blue line in Fig.\ref{fig:4a} indicates the 3D closed elliptical contour, which is the path followed by the quadrotor.
For that, our objective is to conduct an affine transformation, which involves the transformation of a point from a global stationary reference frame $\mathcal{F'}$ to a quadrotor frame $\mathcal{F}$. An affine transformation is a linear mapping that can preserve straight lines,  points and planes but does not necessarily preserve angles and Euclidean distances.

The theory of Affine transformation is as follows:

Consider a three-dimensional domain with a stationary coordinate system $\mathcal{F}'$ as shown in Fig.~\ref{fig:4a}. In this coordinate system, a closed 3D contour $C$, which is an ellipse, is defined. The origin $O$ of a translating coordinate system $\mathcal{F}$ slides on this contour and performs a periodic motion. The relative motion of this system is described as:

\begin{equation}
\begin{cases}
\vec\!{r}'_s[n] = R[n] \, \vec{r}_{s/o}[n] + \vec\!{r}'_o[n]
\label{eq:relative_motion}
\end{cases}
\end{equation}

where \! $\vec\!{r}'_s[n]$ and $\vec{r}_{s/o}[n]$ are the position vectors of a trajectory point in the fixed and moving frames respectively at discrete time step $n$, and \! $\vec\!{r}'_o[n]$ is the position vector of the origin $O$ of the moving frame with respect to the fixed frame at the same discrete time step. The rotation matrix $R[n]$ rotates the moving frame $\mathcal{F}$ to the fixed frame $\mathcal{F}'$ at discrete time step $n$. It should be noted that in the paper, the parameters with a prime $'$ refer to those defined with respect to the fixed coordinate frame, while the others are defined with respect to the moving frame.

For expressing the relative motion equation Eq.\ref{eq:relative_motion} in an affine matrix form, we use the rotation matrix and the translational vector as a matrix. This provides the ability to express the transformation between coordinate systems in a unified matrix format and is also can be computed efficiently. For that, we describe the coordinates, \((x, y, z)\) of a point with respect to the moving frame $\mathcal{F}$ and given translational vector \(\mathbf{t}_{n}\), which describes the origin of the moving frame with respect to the fixed frame as shown in Fig.\ref{fig:4b}. The coordinates  (\(x'\), \(y'\), \(z'\)) of the point with respect to fixed frame $\mathcal{F}'$ can be determined in affine form as follows:

\begin{equation}
\centering
\begin{bmatrix}
x'_{n} \\
y'_{n} \\
z'_{n} \\
1
\end{bmatrix}
=
\begin{bmatrix}
[R]_n & \mathbf{[{t}_n]}^T  \\
0 & 1
\end{bmatrix}
\begin{bmatrix}
x_n \\
y_n \\
z_n \\
1
\label{eq:Affine_matrix}
\end{bmatrix}
\end{equation}
The rotation matrix is first applied to align the coordinates from the moving frame to the global frame during the transformation process and expressed in Eq.\ref{Rot_eqn}
\begin{equation}
\centering
\left([R]_{ij}\right)_n = \mathbf{e}_i' \cdot \mathbf{e}_j, \quad i, j = 1, 2, 3
\label{Rot_eqn}
\end{equation}

where \( \mathbf{e}_1', \mathbf{e}_2', \) and \( \mathbf{e}_3' \) are the unit vectors of the global frame, and \( \mathbf{e}_1, \mathbf{e}_2, \) and \( \mathbf{e}_3 \) are the unit vectors of the moving frame.
The application of the  Euclidean affine transformation matrix Eq.\ref{eq:Affine_matrix} for transforming hyperchaotic points from a fixed frame into a desired moving frame ensures that the hyperchaotic properties of the system are preserved. This is due to the fact that this affine transformation defined by Eq.\ref{eq:Affine_matrix} does not alter the essential characteristics of the chaotic systems, such as boundedness, determinism, and the divergence of nearby trajectories\cite{39}. In other words, this intrinsic hyperchaotic nature of the points remains unchanged even after affine transformation.

\section{Methodology}
\label{methodology}

\begin{figure}[ht!]
    \centering

    \begin{subfigure}[t]{0.7\textwidth}
        \centering
        \includegraphics[width=\textwidth]{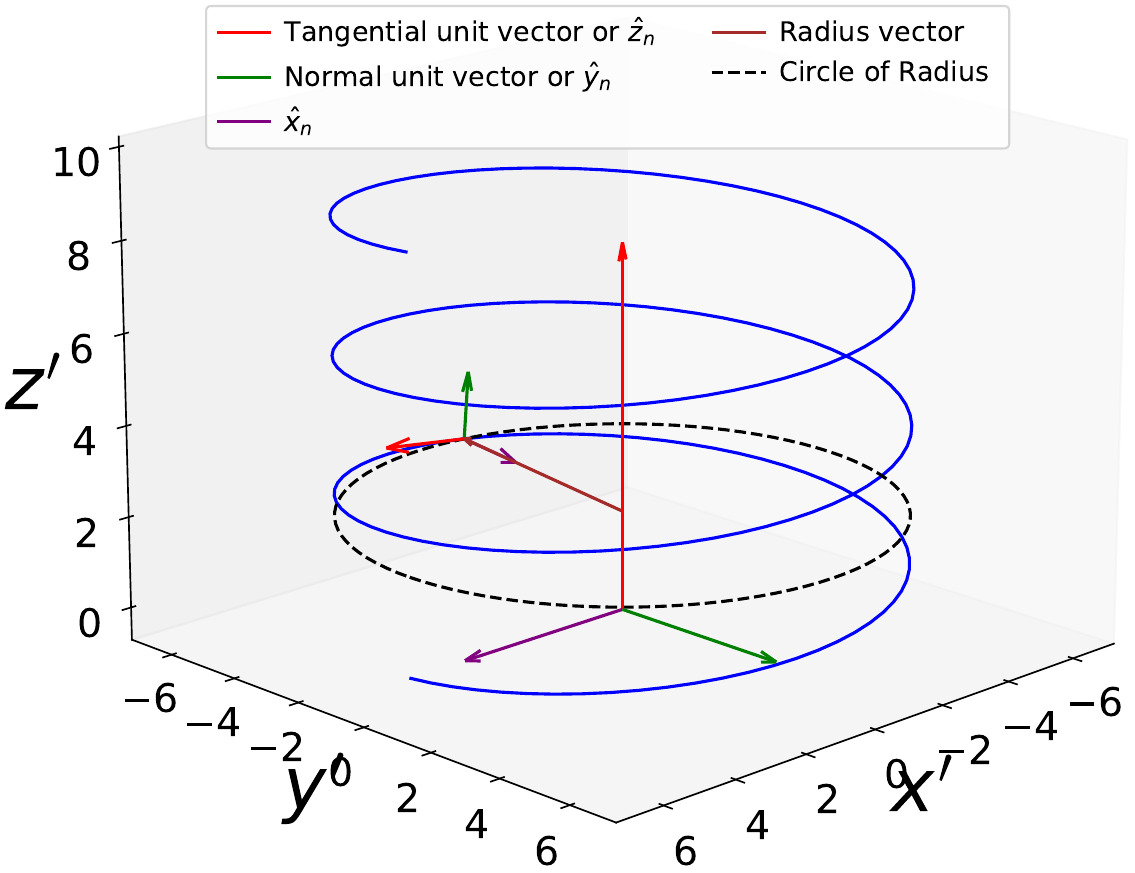}  
        \caption{}
        \label{fig:5a}
    \end{subfigure}
    \vskip\baselineskip
    \begin{subfigure}[t]{0.47\textwidth}
        \centering
        \includegraphics[width=\textwidth]{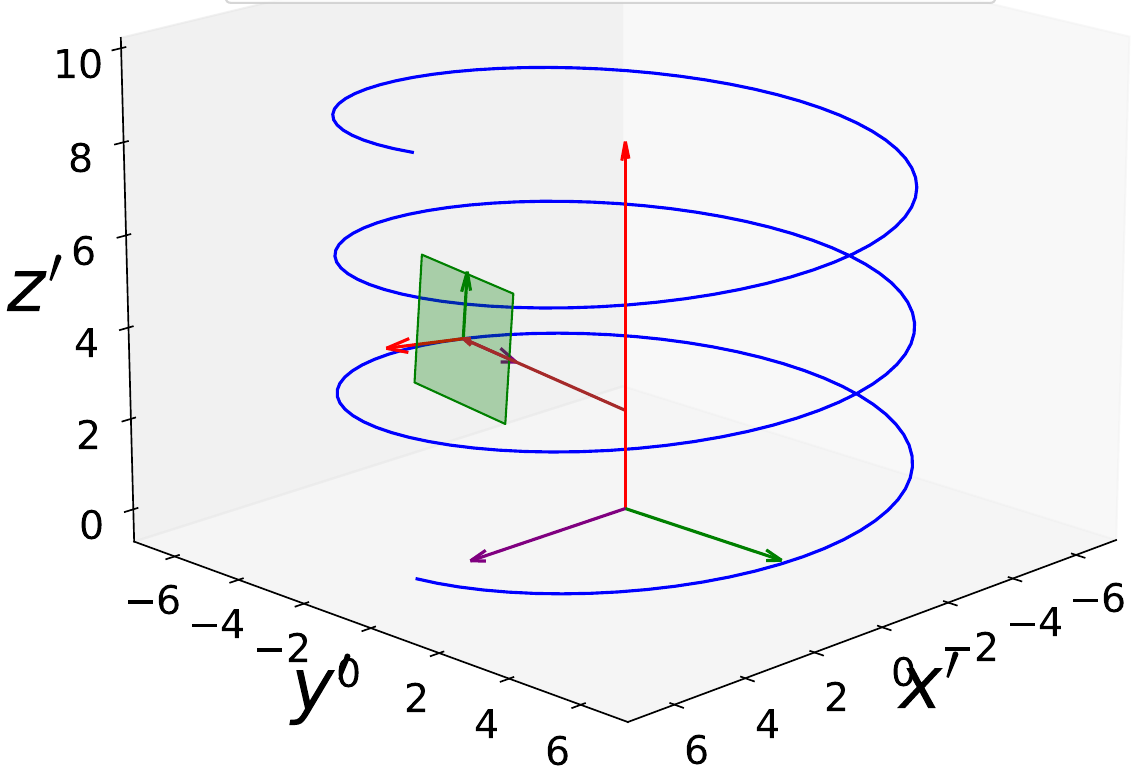}  
        \caption{}
        \label{fig:5b}
    \end{subfigure}
    \hfill
    \begin{subfigure}[t]{0.47\textwidth}
        \centering
        \includegraphics[width=\textwidth]{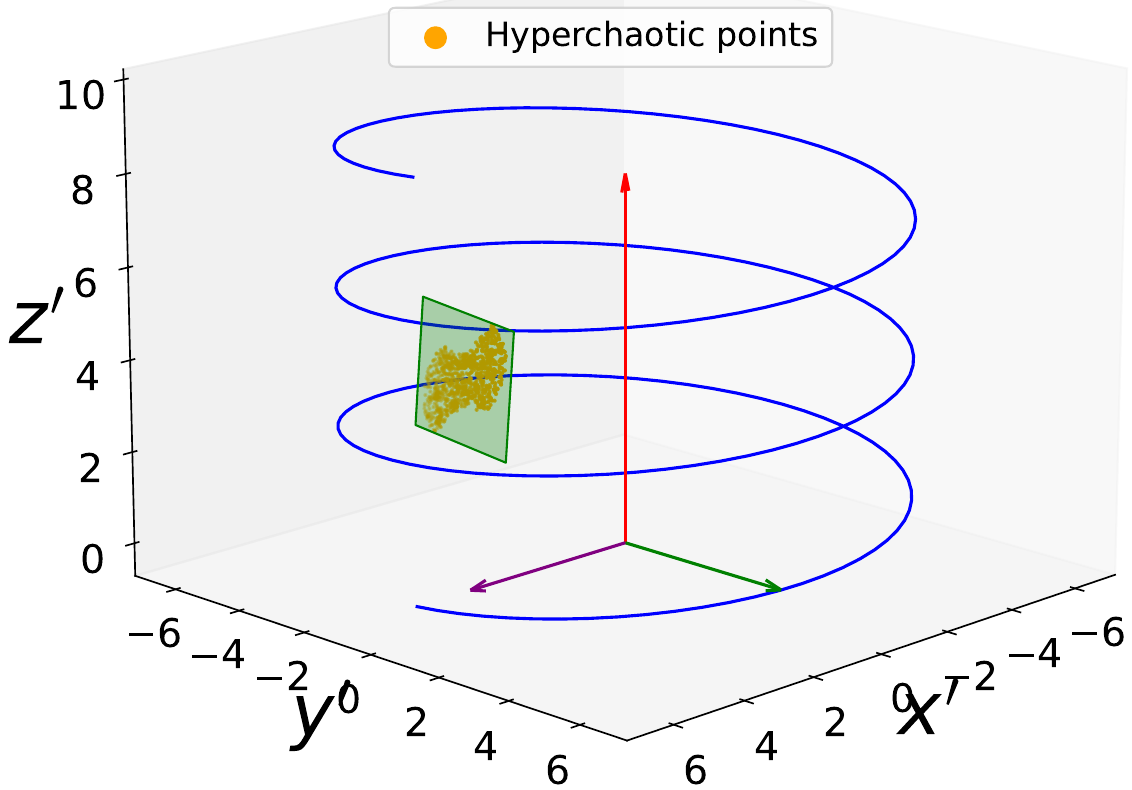}  
        \caption{}
        \label{fig:5c}
    \end{subfigure}
    \caption{Discretization and plotting of $n^{th}$ moving frames along  helical contour. \textbf{a)} The moving coordinates at the $n^{th}$ point on the contour. \textbf{b)} Plot of xy-plane at the $n^{th}$ point on the contour.
    \textbf{c)} Plotting  2D NDMH map on the plane of the $n^{th}$ moving frame. }
    \label{fig:moving_coordinates/plane/NDMH_map}
\end{figure}
The comprehensive overview of the proposed algorithm or method is as follows:

A plane holding a two-dimensional hyperchaotic NDMH map slides along an open helical contour, progresses vertically upward around a building, and acts as a guiding line for a quadrotor. The normal vector of the plane is always kept parallel to the tangential vector of the contour at every discrete point. After generating a sequence of points using an NDMH map across successive planes, like the nth iterated point from the system is on the nth plane, joining them with a curve yields the desired trajectory.

This paper presents an open helical contour that progresses incrementally around vertical structures like buildings according to a pitch value. A pitch is defined as the distance between two consecutive turns in a helix. Parametric equations Eq.\ref{eq:contour_eqn} are used to plot the contour of the helix, using parameters like initial radius $r$, radius growth rate $g$, and pitch as $p$. The initial radius sets the starting size of the helix, while the radius growth rate describes how the radius may increase as in Fig.\ref{fig:b} or decrease as in Fig.\ref{fig:c} along the helix. And if the radius growth rate is 0, then the contour will be the cylindrical helix as in Fig.\ref{fig:a}. These parameters are used as follows to define the contour of the helix.

\begin{equation}
\begin{cases}
x'(t) = \left(\text{r} + \text{g} \* t\right) \cos(t) \\
y'(t) = \left(\text{r} + \text{g}  \* t\right) \sin(t) \\
z'(t) = \frac{\text{p}  \* t}{2\pi}

\label{eq:contour_eqn}
\end{cases}
\end{equation}
To determine the tangent vectors along the helical contour, the first derivative of the parametric equations Eq.\ref{eq:contour_eqn} with respect to t are taken. The derivatives below give the components of the tangent vector at any point along the helix.

\begin{equation}
\begin{cases}
\frac{dx'(t)}{dt} &= -\left(\text{r} + \text{g} \* t\right) \sin(t) + \text{g} \* \cos(t) \* t \\
\frac{dy'(t)}{dt} &= \left(\text{r} + \text{g} \* t\right) \cos(t) + \text{g}\*t \* \sin(t)  \\
\frac{dz'(t)}{dt} &= \frac{\text{p}}{2\pi}
\label{eq:tangent_eqns}
\end{cases}
\end{equation}

 The contour is then discretized to account for accurate computations and simulations. This discretization is necessary to adapt the kinematic relative motion equations Eq.\ref{eq:relative_motion} for an $n^{th}$ point on the helical contour. For that, a parameter  \textit{steps} is added to the algorithm to generate a specified number of equidistant points along the open helical contour. Then, the next goal is to plot moving frames at those discrete points. Fig.\ref{fig:5a} depicts the $n^{th}$ point on the general helical contour, represented by coordinates $(x_n', y_n', z_n')$, and a radius vector is drawn to a point $(0, 0, z_n')$ on the global $z'$ axis from this particular point, which shown as a brown line in Fig.\ref{fig:5a}. This radius vector corresponds to the circle passing through the $n^{th}$ point on the contour, as depicted by the dotted black line. In this figure, the global z'-axis has been extended for clarity to show where the radius vector actually connects. And it is considered that the quadrotor moves from the bottom toward the top in an anticlockwise direction. So the direction of the tangent follows this motion, as illustrated in Fig.\ref{fig:5a}. The normal vector is obtained by taking the cross product of the tangent vector and the radius vector, as given below.

\begin{equation}
\vec{ N}(n) =\vec{t}(n)  \times \vec{r}(n)
\label{eq:norm_vec}
\end{equation}

Where \(\vec{t}(n)\)  is calculated based on the Eq.\ref{eq:tangent_eqns} above, at an $n^{th}$ point on contour and \(\vec{r}(n)\) is the radius vector from the $n^{th}$ point on contour. Note that Eq.\ref{eq:norm_vec} is an adapted equation due to the direction of the radius vector taken from the $n^{th}$ point on the contour to the global z' axis.

The unit vectors of the normal vector, radius vector and tangent vector are denoted as $\hat{y}_n$, $\hat{x}_n$, and $\hat{z}_n$ respectively, which represent the axes of moving frame at the $n^{th}$ point on the contour is shown in Fig.\ref{fig:5a}. For an arbitrary contour curve, which is not the defined open helical contour, the above mentioned method for finding the moving coordinate system is not valid. As pointed out by  Gohari et al.\cite{26} in these instances, $\vec\! {z}_n$ is considered to be the vector pointing towards the next point from current point in the contour, and $\vec\!{z}_0$ is the position vector from the origin of the global frame to the first point in the contour. Then $\hat{z}_n, \hat{y}_n, \text {and} \hat{x}_n$ are written as.

\begin{equation}
\begin{cases}
\hat{z}_n &= \frac{\vec\!{p}'_{n+1} - \vec\!{p}'_n}{\left\lVert \vec\!{p}'_{n+1} - \vec\!{p}'_n \right\rVert}, \\
\hat{y}_n &= \frac{\hat{z}_0 \times \hat{z}_n}{\left\lVert \hat{z}_0 \times \hat{z}_n \right\rVert}, \\
\hat{x}_n &= \hat{y}_n \times \hat{z}_n,

\end{cases}
\end{equation}

where  $\vec\!p'_n$ is the position vector of the $n^{th}$ point on the contour.
In the above case, one constraint is that when part of the guiding line is almost parallel to the reference unit vector \(\hat{z}_0\), the $\hat{y}_n$ in that case cannot be defined since \(\hat{z}_0 \times \hat{z}_n = 0\). The \(y\)-axis in such a case can be selected though as \(\hat{y}_n = \frac{\hat{z}_{n-1} \times \hat{z}_n}{\|\hat{z}_{n-1} \times \hat{z}_n\|}\). However, in the proposed method, this constraint does not apply, as the radius vector is used to plot the normal vector. Because the radius vector will always be perpendicular to the tangential vector.

The kinematic relative motion Eq.\ref{eq:relative_motion} is used for transforming points in 3D coordinates. However, our approach requires transformation only in two dimensions. So the Eq.\ref{eq:relative_motion} is reduced as below for that purpose.

\begin{figure}[!ht]
    \centering
    
    \begin{subfigure}[t]{0.5\textwidth}
        \centering
        \includegraphics[width=\textwidth]{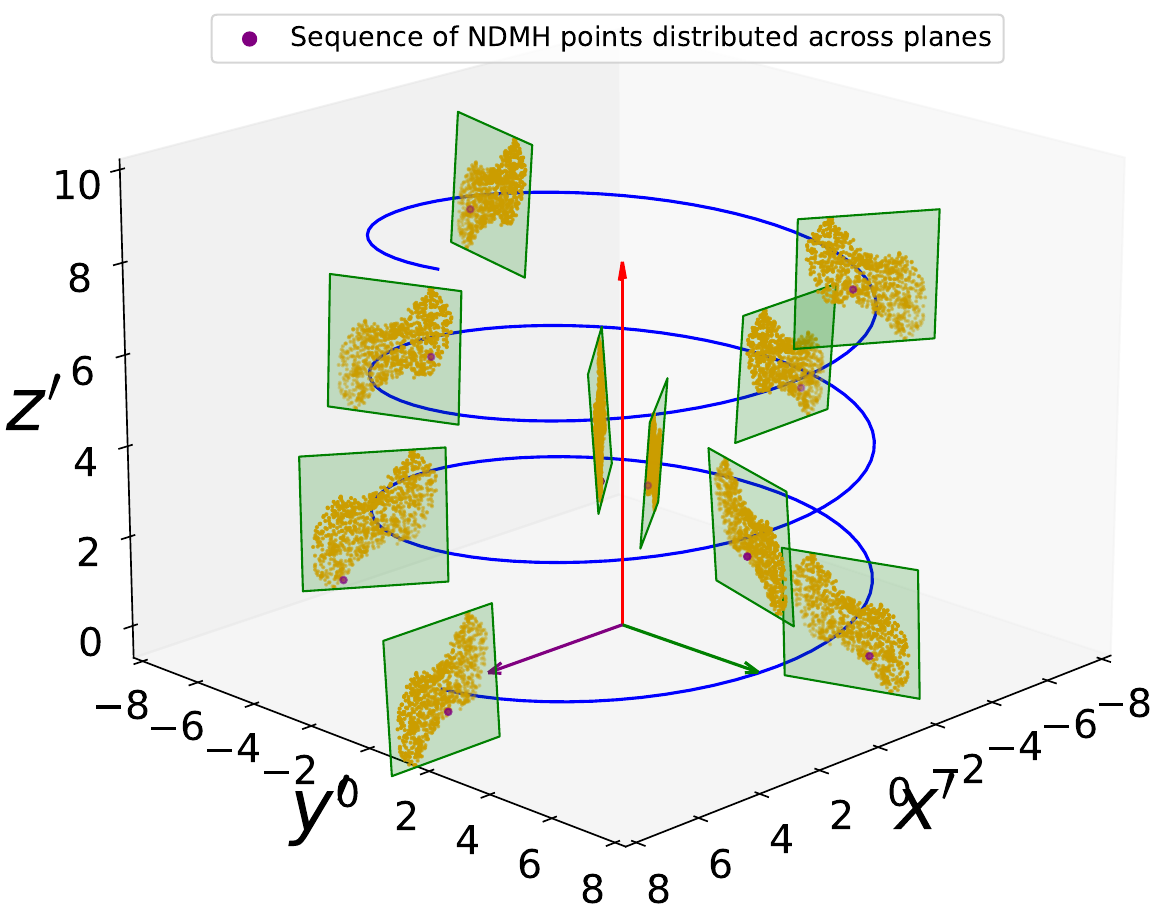}  
        \caption{}
        \label{fig:6a}
    \end{subfigure}
    
    \vskip\baselineskip
    \begin{subfigure}[t]{0.5\textwidth}
        \centering
        \includegraphics[width=\textwidth]{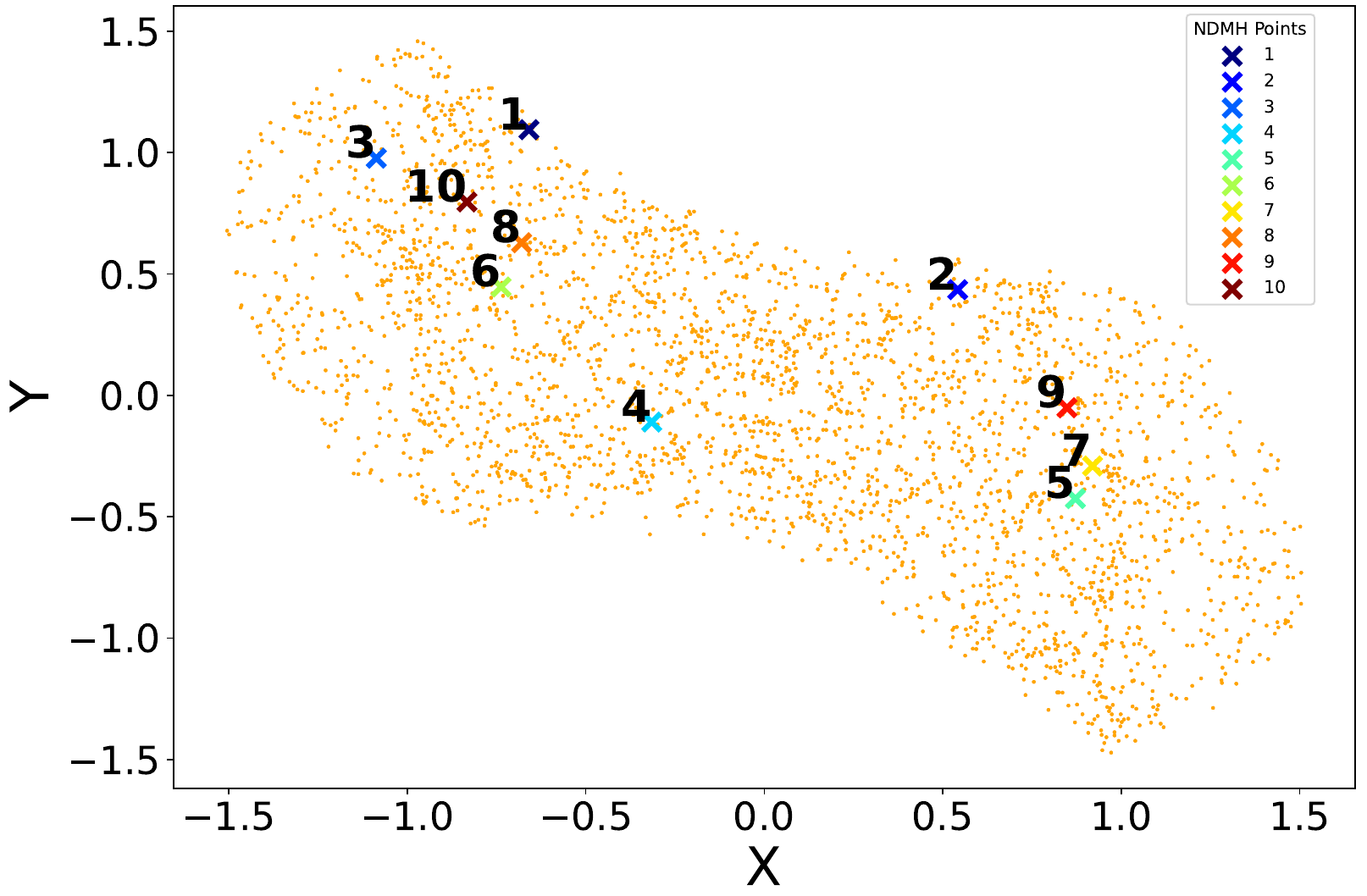}  
        \caption{}
        \label{fig:6b}
    \end{subfigure}
    
    \caption{Visualization of the sequence of points from the hyperchaotic map.\textbf{a)} Sequence of points plotted across successive planes on all xy planes of the moving frames, represented as purple dots.\textbf{b)} Inverse transform of the taken sequence of points onto a common plane.}
    \label{fig:6}
\end{figure}

\begin{figure}[!ht]
    \centering
    \includegraphics[width=0.4\textwidth]{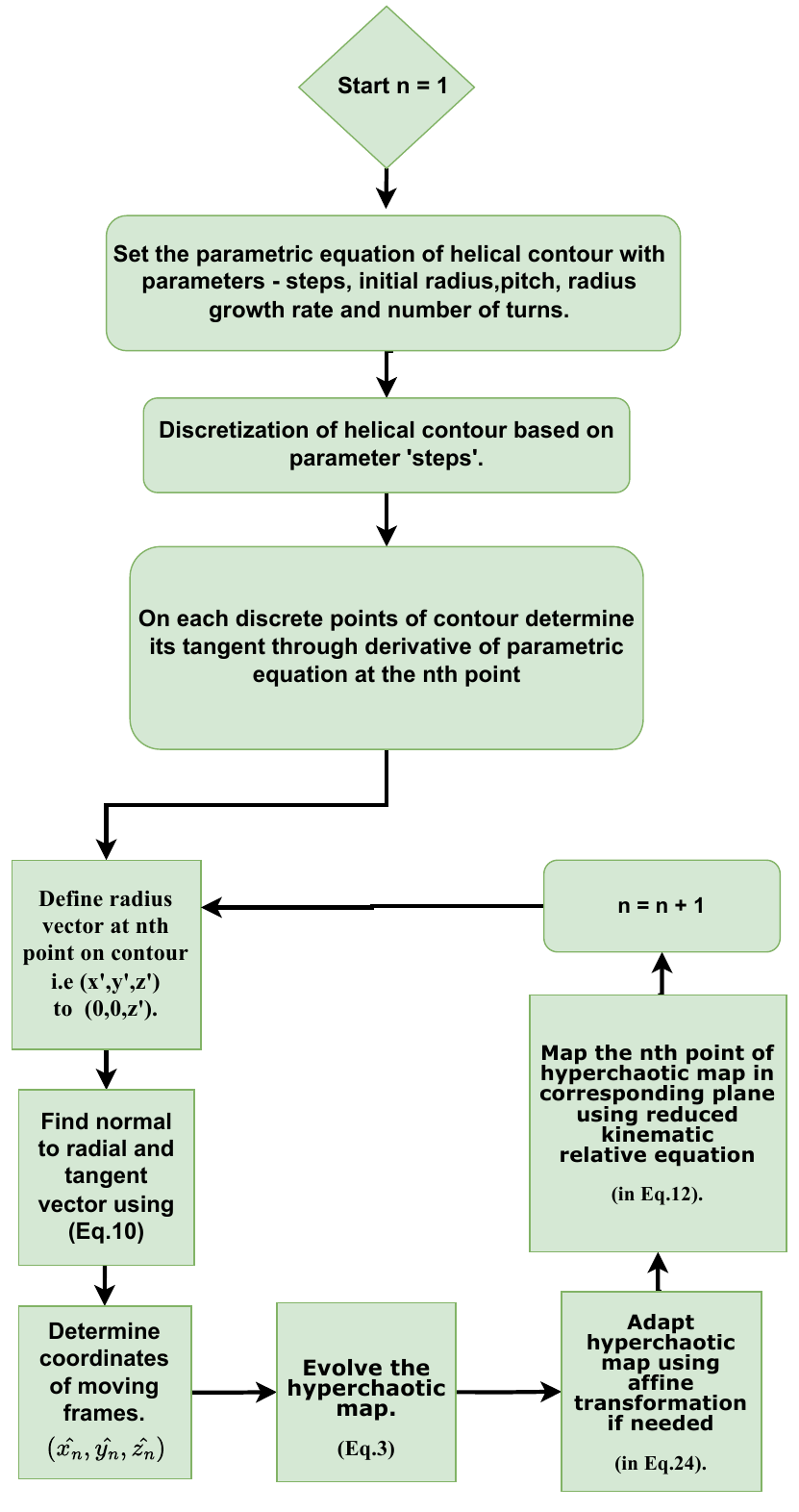}
    \caption{Flowchart of the proposed algorithm for generating hyperchaotic trajectories in 3D space.}

    \label{fig:Flowchart}
\end{figure}

\begin{equation}
\vec{P}' = \vec\!{p}'_n + x (\ \vec\!{r}(n)) + y \ \left(\vec\!{r}(n) \times \vec\!{t}(n)\right)
\label{eq:red_relative_eqn}
\end{equation}

where   $\vec{P}'$ vector represents the position vector of a transformed point with respect to the fixed global frame, and $\vec \!p'_n$ is the position vector of the $n^{th}$ point on the contour with respect to the origin of the global frame. This equation is meant to plot a point in the 2D xy plane of the moving frame relative to the global frame. Through this reduced affine transformation Eq.\ref{eq:red_relative_eqn}, any point can be plotted on the xy plane of any $n^{th}$ moving frame while preserving its geometric affinity.
Strogatz et al.\cite{39} validated that kinematic relative motion in Eq.\ref{eq:relative_motion} is affine. By sequentially applying the hyperchaotic points through this Eq.\ref{eq:red_relative_eqn}, the 2D NDMH map is effectively plotted on the $n^{th}$ moving frame along the contour, which is shown in Fig.\ref{fig:5c}. Prior to this, Eq.\ref{eq:red_relative_eqn} is utilized to plot the $xy$ plane of specific widths on the $n^{th}$ moving frame by taking coordinates of vertices and visualised in Fig.\ref{fig:5b}. A desired trajectory is obtained by mapping a sequence of points along this series of planes, like $n^{th}$ sequential point on the nth plane along the contour, and connecting them with a smooth or linear curve.

The verification of whether the sequence of points are part of a hyperchaotic map was conducted through visualization. For that, the NDMH map was plotted on all $xy$ planes of the moving frames, and the sequence of points are plotted across successive planes, as in shown Fig.\ref{fig:6a} as purple dots. The number of planes is taken as 10, as determined by the parameter 'steps', which is set to 10. Since the initial iteration starts from 47000, then $47010^{th}$ point of the sequence lies on the last $10^{th}$ plane along the contour. Then, the position coordinates of these sequences of points on all planes were subjected to an inverse transformation of Eq.\ref{eq:red_relative_eqn} to a common plane as shown in Fig.\ref{fig:6b}. And different colours are assigned to each point extracted from the planes. Since the sequence started after 47000 iterations, here in the common plane shown in Fig\ref{fig:6b} has NDMH points plotted till 50000. In this way, it is possible to confirm if they were part of a hyperchaotic attractor or not. Fig.\ref{fig:Flowchart} shows the flowchart of
proposed method.

\section{Quadrotor simulation model}
\label{quadrotor_simulation_model}

\begin{figure*}[htbp]
    \centering
    \begin{subfigure}[b]{0.5\textwidth}
        \centering
        \includegraphics[width=\textwidth]{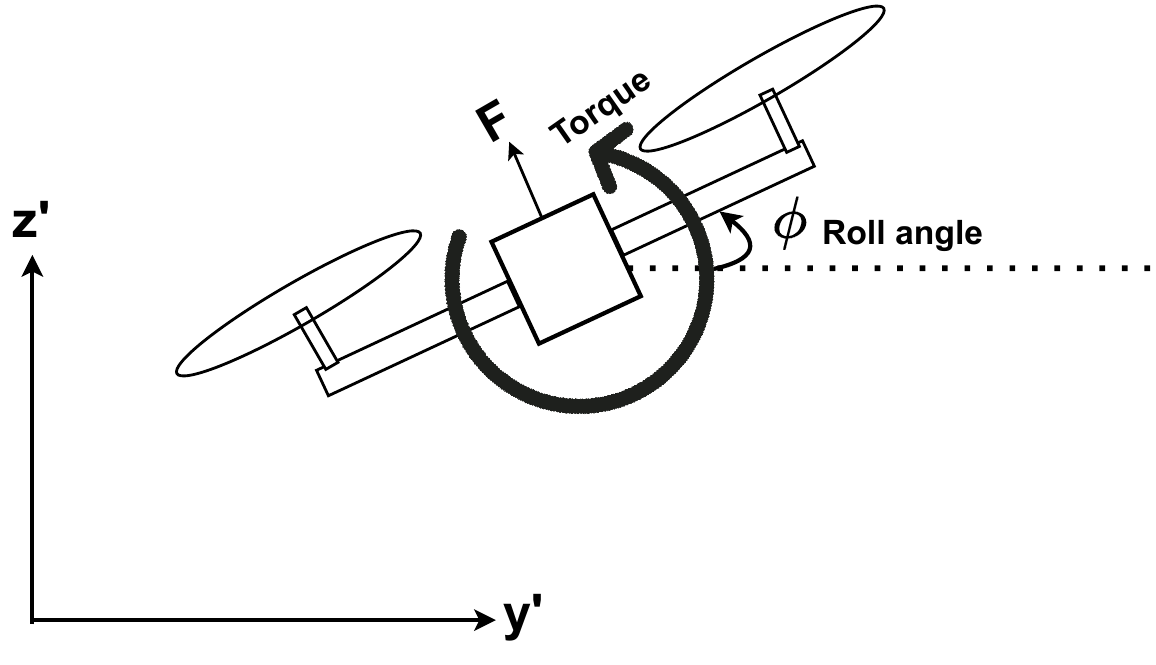}
        \caption{}
        \label{fig:qdr1}
    \end{subfigure}
    \hfill
    \begin{subfigure}[b]{0.44\textwidth}
        \centering
        \includegraphics[width=\textwidth]{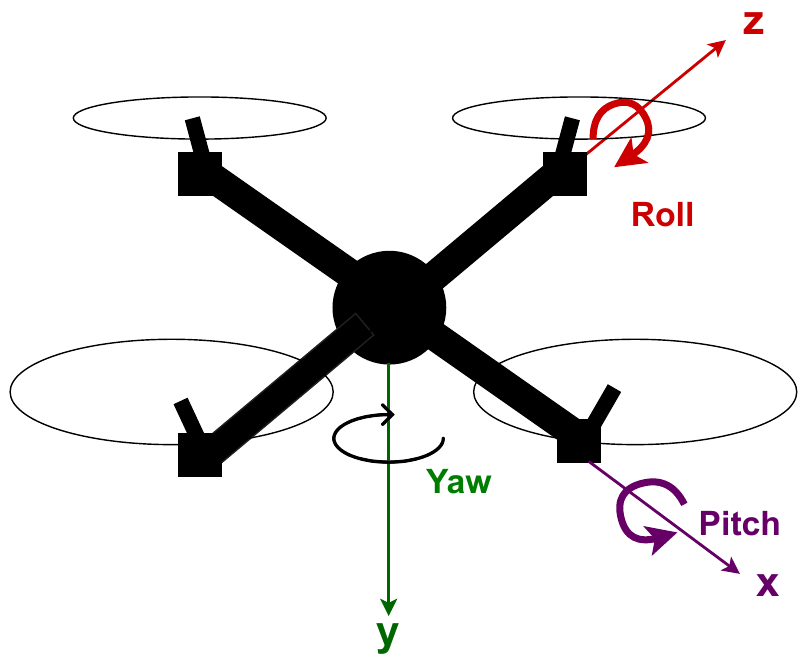}
        \caption{}
        \label{fig:qdr2}
    \end{subfigure}
    \caption{Illustration of quadrotor dynamics and rotation matrices. \textbf{a)} Thrust force during roll motion. \textbf{b)} Basic rotation matrices: roll, pitch and yaw with corresponding axes of rotation.}
    \label{fig:two_images}
\end{figure*}

The quadrotor is a four-rotor UAV. It works based on a set of dynamical equations that describe the motion and control of this quadrotor \cite{59}. Kinematically, the quadrotor's placement in 3D space is modelled by a position vector \(\mathbf{p} = [x, y, z]^T\) , while the orientation is modelled by roll (\(\phi\)), pitch (\(\theta\)), and yaw (\(\psi\)) which is shown in Fig.\ref{fig:qdr2}. The rotation matrix \(\mathbf{R}_q\) transforms the forces and torques from the body frame into the inertial frame and is expressed as:

\begin{equation}
\mathbf{R}_q = \mathbf{R}_qz(\psi) \cdot \mathbf{R}_qy(\theta) \cdot \mathbf{R}_qx(\phi)
\label{Rq}
\end{equation}

\paragraph{Roll (\(\phi\))}: The rotation about the \(x\)-axis is given by:
\begin{equation}
\mathbf{R}_qx(\phi) = \begin{bmatrix}
1 & 0 & 0 \\
0 & \cos \phi & -\sin \phi \\
0 & \sin \phi & \cos \phi
\end{bmatrix}
\label{17}
\end{equation}

\paragraph{Pitch (\(\theta\))}: The rotation about the \(y\)-axis is given by:
\begin{equation}
\mathbf{R}_qy(\theta) = \begin{bmatrix}
\cos \theta & 0 & \sin \theta \\
0 & 1 & 0 \\
-\sin \theta & 0 & \cos \theta
\end{bmatrix}
\label{18}
\end{equation}

\paragraph{Yaw (\(\psi\))}: The rotation about the \(z\)-axis is given by:
\begin{equation}
\mathbf{R}_qz(\psi) = \begin{bmatrix}
\cos \psi & -\sin \psi & 0 \\
\sin \psi & \cos \psi & 0 \\
0 & 0 & 1
\end{bmatrix}
\label{19}
\end{equation}

where \(\mathbf{R}_qx(\phi)\), \(\mathbf{R}_qy(\theta)\), and \(\mathbf{R}_qz(\psi)\), are the basic rotation matrices for roll, pitch, and yaw respectively, which is shown in Fig.\ref{fig:qdr2} \cite{64}. An equal adjustment in the speed of the diagonal motors causes a change in the yaw angle. Conversely, if the rotational speed of one rotor on the roll/pitch axis is altered relative to the corresponding rotor on the same axis, it results in roll or pitch motion.The dynamics associated with translation are due to the thrust force expressed by:

\begin{equation}
\mathbf{F} = m \cdot (g + \mathbf{a})
\end{equation}

In that context, $m$ stands for the mass of the quadrotor, $g$ is the acceleration of gravity, and $\mathbf{a}$ is the acceleration that would be desired. Specifically, thrust force can be written as:

\begin{align}
\mathbf{F} = m \cdot \Big(&g + \text{z\_acc} \nonumber \\
&+ K_{p_z} \cdot (\text{z\_pos} - z) \nonumber \\
&+ K_{d_z} \cdot (\text{z\_vel} - \dot{z})\Big)
\label{eq:thrust}
\end{align}

where \(\text{z\_acc}\) is the desired vertical acceleration, \\  \(\text{z\_pos}\) is the desired vertical position,\(\text{z\_vel}\)  is the desired vertical velocity  and\(K_{p_z}\) and \(K_{d_z}\) are the proportional and derivative gains, respectively.During rolling, the thrust force shown in Fig.~\ref{fig:qdr1} highlights the direction of it with respect to the tilt of the quadrotor.  The rotation dynamics are given by:

\begin{equation}
\mathbf{\tau} = \mathbf{I} \cdot \boldsymbol{\alpha} + \boldsymbol{\omega} \times (\mathbf{I} \cdot \boldsymbol{\omega})
\end{equation}

where \(\mathbf{I}\) is the moment of inertia matrix, \(\boldsymbol{\alpha}\) is the angular acceleration, and \(\boldsymbol{\omega}\) is the angular velocity vector.Torque \( \tau \) is shown in Fig.~\ref{fig:qdr1} as the rotational force applied to the quadrotor, making it rotate around its rolling axes. Control inputs are applied through thrust control, a method that works on the vertical movement of the object as in Eq.\ref{eq:thrust} and torque control, which controls the roll, pitch, and yaw of the object as in Eqns.\ref{17}, Eqns.\ref{18}, Eqns.\ref{19} respectively. Dynamic principles and control strategies are set in place to ensure stability and maintain the desired trajectories of flights.

The full transformation matrix \(\mathbf{T}\) combines rotation and translation as given below.

\begin{equation}
\mathbf{T} = \begin{bmatrix}
\mathbf{R}_q & \mathbf{p} \\
0 & 1
\end{bmatrix}
\end{equation}

where \(\mathbf{R}_q\) is the \(3 \times 3\) rotation matrix calculated based on Eq.\ref{Rq} and \(\mathbf{p} = \begin{bmatrix} x & y & z \end{bmatrix}^T\) is the position vector. The last row \([0\;1]\) is added to make \(\mathbf{T}\) a homogeneous transformation matrix.

The positions of the quadrotor's four rotors in its local frame are defined as:
\begin{equation}
\begin{aligned}
\mathbf{p}_1 &= \begin{bmatrix}
\frac{\text{size}}{2} \\
0 \\
0 \\
1
\end{bmatrix}, \quad
\mathbf{p}_2 = \begin{bmatrix}
-\frac{\text{size}}{2} \\
0 \\
0 \\
1
\end{bmatrix}, \\
\mathbf{p}_3 &= \begin{bmatrix}
0 \\
\frac{\text{size}}{2} \\
0 \\
1
\end{bmatrix}, \quad
\mathbf{p}_4 = \begin{bmatrix}
0 \\
-\frac{\text{size}}{2} \\
0 \\
1
\end{bmatrix}
\end{aligned}
\label{eq:points}
\end{equation}

where \(\text{size}\) denotes the distance between the centre of the quadrotor and each rotor. To find out the positions of the rotors in the global frame, these local frame coordinates are transformed by utilizing the transformation matrix \(T\). This transformation maps the rotor positions from the local coordinate system to the global coordinate system, considering the quadrotor’s orientation and position.

In quadrotor trajectory planning, quintic polynomials are used to generate smooth transitions between waypoints by defining the precise position, velocity, acceleration profiles and $T_P$ trajectory duration. These polynomials \( p(t) \) are described by the general form
\begin{equation}
p(t) = a_0 + a_1 t + a_2 t^2 + a_3 t^3 + a_4 t^4 + a_5 t^5,
\end{equation}
where the coefficients $a_0$ through $a_5$ are determined by solving the linear equations based on the boundary conditions given for the initial and final positions, velocities, and accelerations. Along with the initial and final positions of waypoints given, we will define a parameter  $T_P$ for the trajectory duration. It influences the quadrotor motion, whereby a larger $T_P$ results in smoother but more slowly joining paths, and a smaller $T_P$ gives faster transitions that provide abrupt motion or instability. In our case of the simulation, a quintic polynomial approach for the quadrotor trajectory is calculated under the initial and final position and $T_P$. The coefficients $\ensuremath{a}$ ensure that the path satisfies all of the constraints imposed upon it, such as that to provide smoothness from start and end. Once obtained, these coefficients are used to calculate the position, velocity, and acceleration profiles at any given time, enabling the quadrotor to follow a smooth path. The position, velocity, and acceleration profiles provide the necessary references for the control algorithms to adjust the quadrotor's thrust and attitude to follow the trajectory accurately. Additionally, a PID control system is employed to fine-tune the quadrotor's response and eliminate error accumulation while following the given trajectories. In this paper, the waypoints that fed into the quintic polynomials are hyperchaotic trajectories generated by the algorithm discussed in the previous section. The sequence hyperchaotic points are plotted across consecutive planes along the contour, where the $n$-th point lies on the $n$-th plane, and the $(n+1)$-th point lies on the $(n+1)$-th plane. These $ n$-th and $(n+1)$-th points are used as waypoints in the quintic polynomial algorithm. This process is further iterated to create the entire quadrotor trajectory, with the time parameter $T_P$ determining the duration of the trajectory between consecutive points.

\begin{figure*}[ht!]
    \centering
    \begin{subfigure}[t]{0.49\textwidth}
        \centering
        \includegraphics[width=\textwidth]{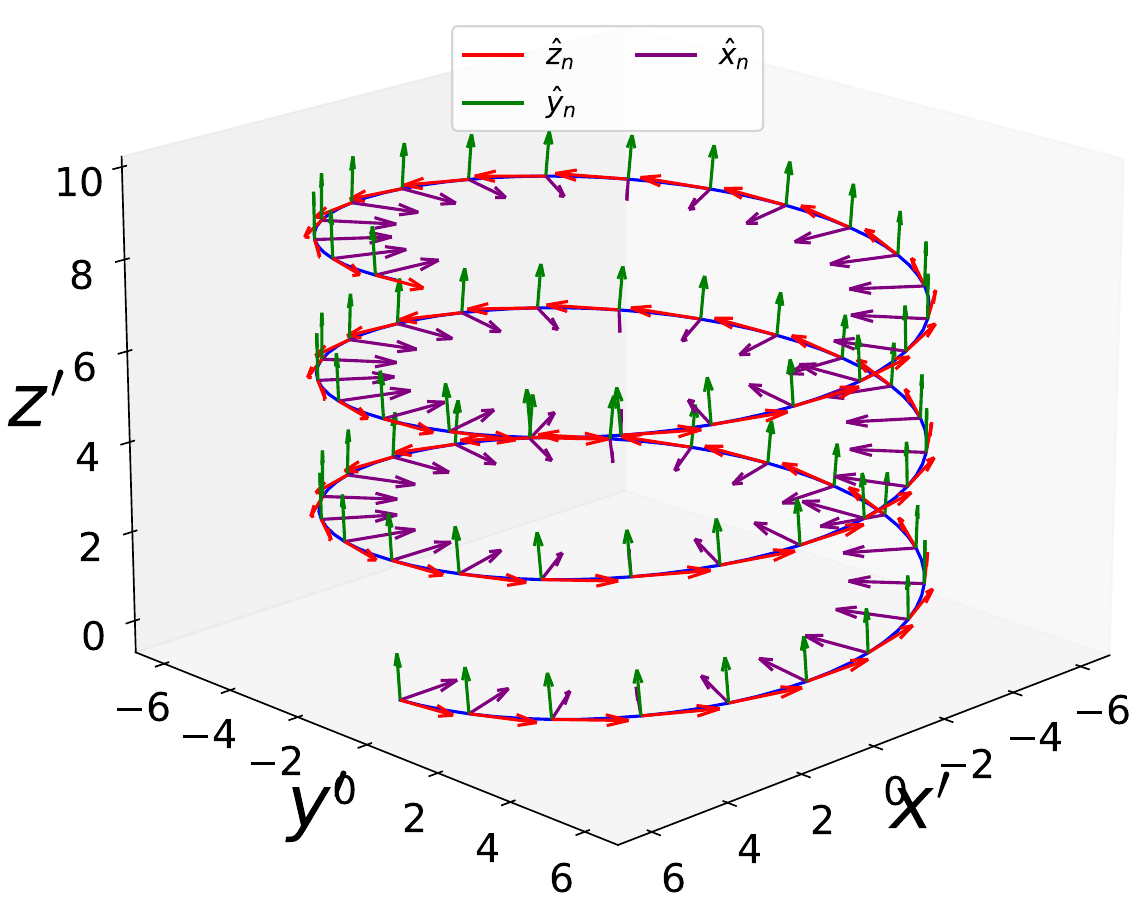}  
        \caption{}
        \label{fig:image1}
    \end{subfigure}
    \hspace{5pt} 
    \begin{subfigure}[t]{0.47\textwidth}
        \centering
        \includegraphics[width=\textwidth]{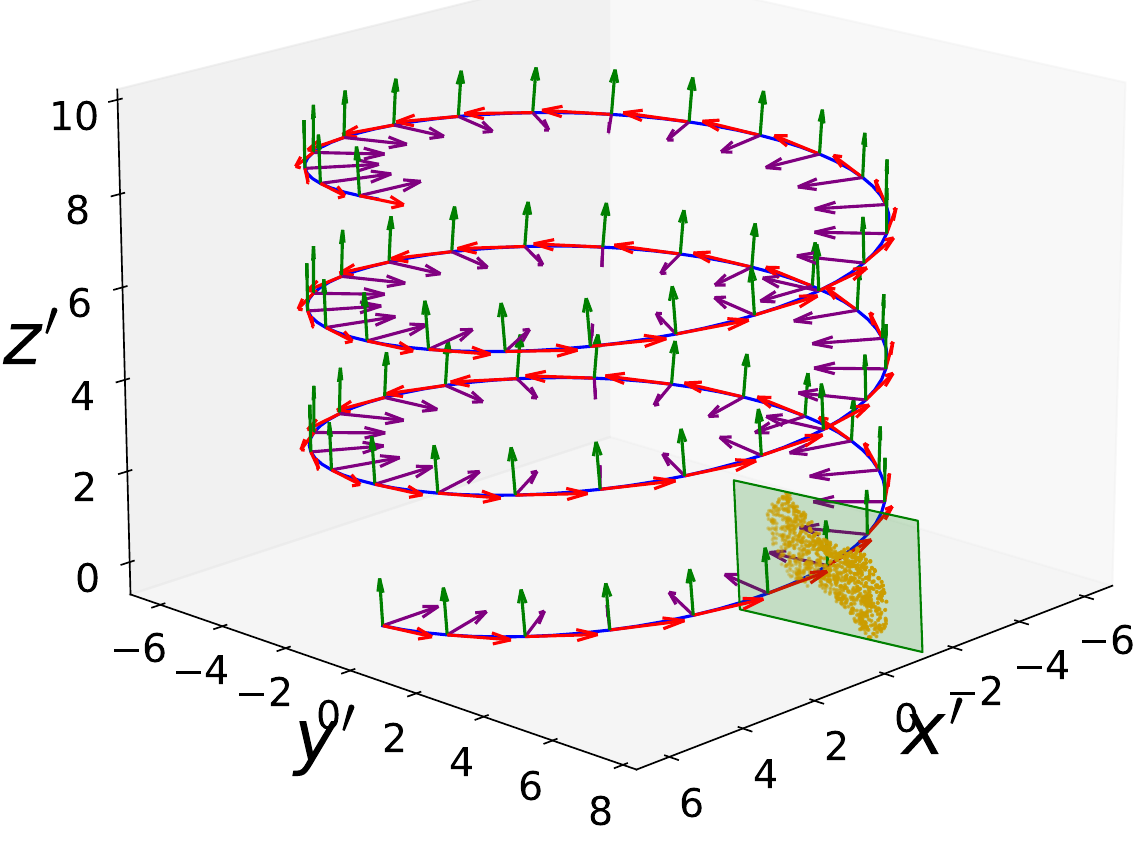}  
        \caption{}
        \label{fig:image2}
    \end{subfigure}
    
    \vskip\baselineskip
    \begin{subfigure}[t]{0.49\textwidth}
        \centering
        \includegraphics[width=\textwidth]{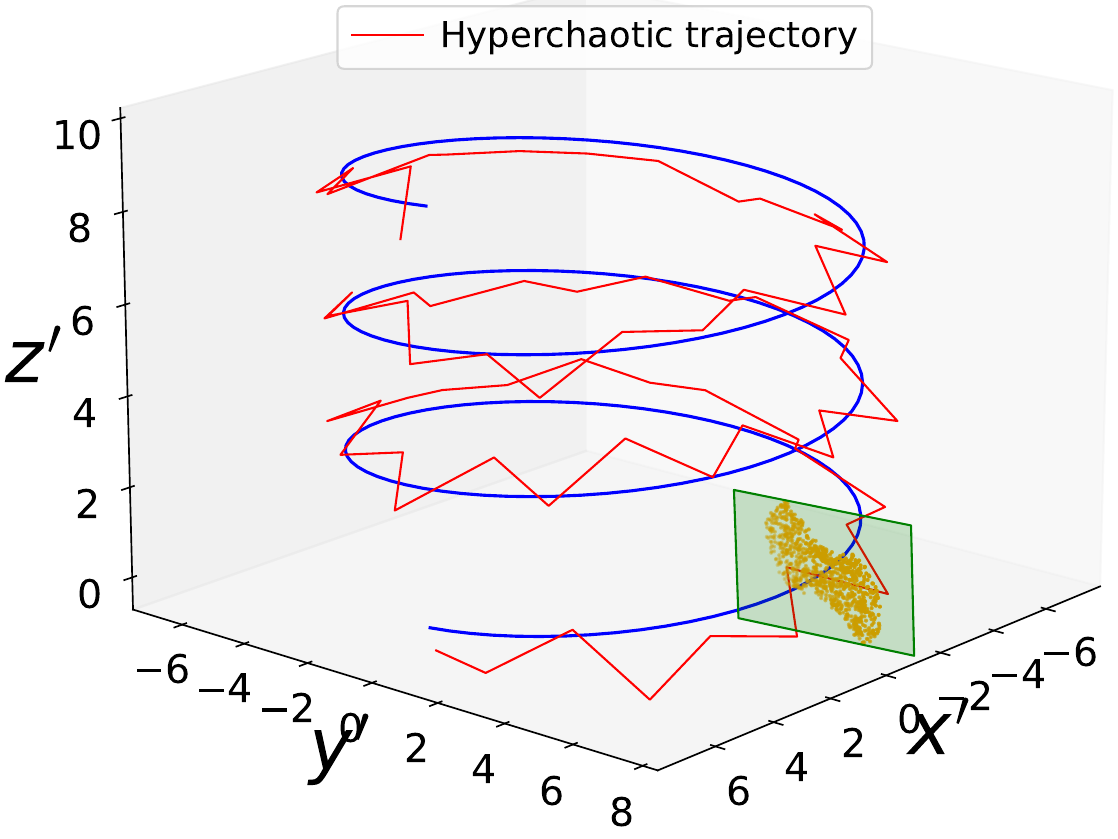}  
        \caption{}
        \label{fig:image3}
    \end{subfigure}
    \hspace{5pt} 
    \begin{subfigure}[t]{0.47\textwidth}
        \centering
        \includegraphics[width=\textwidth]{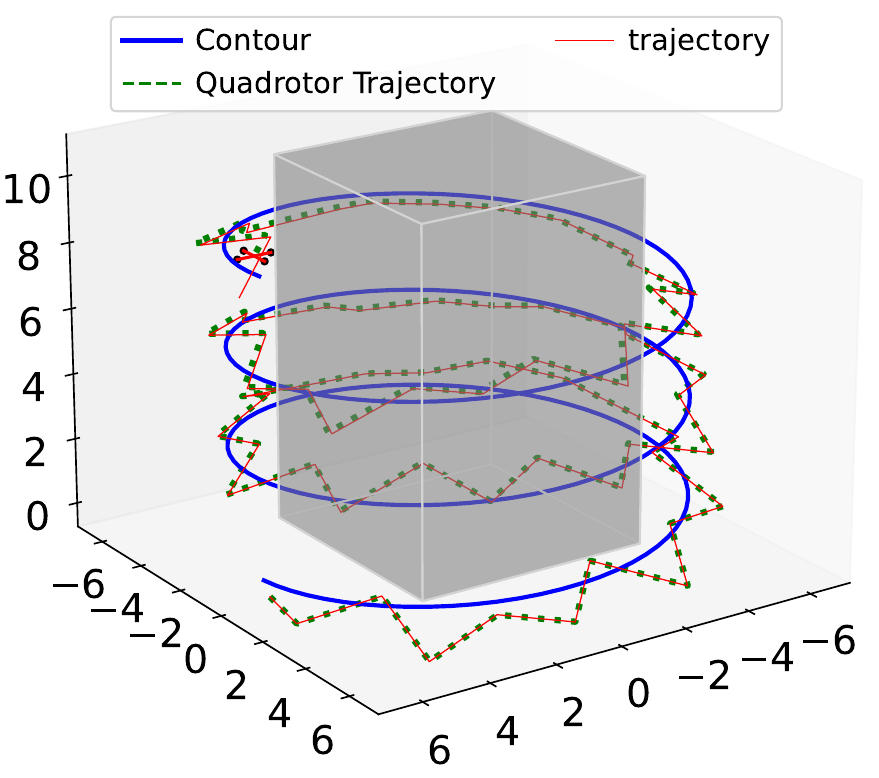}  
        \caption{}
        \label{fig:image4}
    \end{subfigure}
    
    \caption{Quadrotor trajectory illustration: \textbf{a)} Moving coordinates in 60 equidistant points along contour\textbf{b)} Mapping of 2D NDMH map on the seventh moving frame \text{c)} Plot of showing hyperchaotic trajectory as red line \textbf{d)} Plot of simulated quadrotor trajectory as green dotted line }
    \label{fig:all_images}
\end{figure*}
\section{Results and simulations}
\label{results}

In the initial phase of this study, a 3D open general helical contour was defined utilizing parametric equations Eq.\ref{eq:contour_eqn}  with a pitch of 3 and three turns. In order to avoid congestion, the step parameter of 60 was chosen, which makes such numbers of equidistant moving coordinates plotted along the helical contour, as illustrated in Fig\ref{fig:image1}. On each plane within the moving frame, the 2D NDMH  map was mapped using the reduced kinematic relative equation provided in  Eq.\ref{eq:red_relative_eqn}, which is shown only in the seventh plane for better observation in Fig.\ref{fig:image2} as given above. Note that labels for moving frames, contours, and planes are consistent throughout the paper, with the same colours and labels used uniformly. And started the iteration from 49000 in order to get a higher divergence of the sequence of the points and generated hyperchaotic trajectory is shown in Fig.\ref{fig:image3} as red lines. Fig.\ref{fig:image4} shows the Quadrotor trajectory, simulated corresponding to the given conditions with trajectory duration of $T_P$ = 60 and quadrotor mass of 200g. The green dotted line shows the path traced by a quadrotor, and instead of a building, it shows a grey rectangular box, which is surrounded by a helical contour. The simulation was done utilizing PythonRobotics\cite{63} library that offers an updated simulation environment for quadrotor path planning and control.

Scaling is a special case of affine transformation characterized by its linearity, where all the coordinates are multiplied by a scalar factor and with no translation involved, as given below

\begin{equation}
\mathbf{x}' = \mathbf{S} \mathbf{x} + \mathbf{b} =\mathbf{S} \mathbf{x}, \text{ where intercept b = 0 } 
\label{eq:scaling}
\end{equation}
ie The hyperchaotic equations of NDMH map is termed as,
\begin{equation}
\begin{bmatrix}
x_{n+1} \\
q_{n+1}
\end{bmatrix}
=
\mathbf{S}
\begin{bmatrix}
 k (q_n^2 - 1) x_n \\
q_n + x_n
\end{bmatrix}
\label{eq:scaled_matrix}
\end{equation}

\begin{figure*}[ht!]
    \centering
    \begin{subfigure}[t]{0.45\textwidth}
        \centering
        \includegraphics[width=\textwidth]{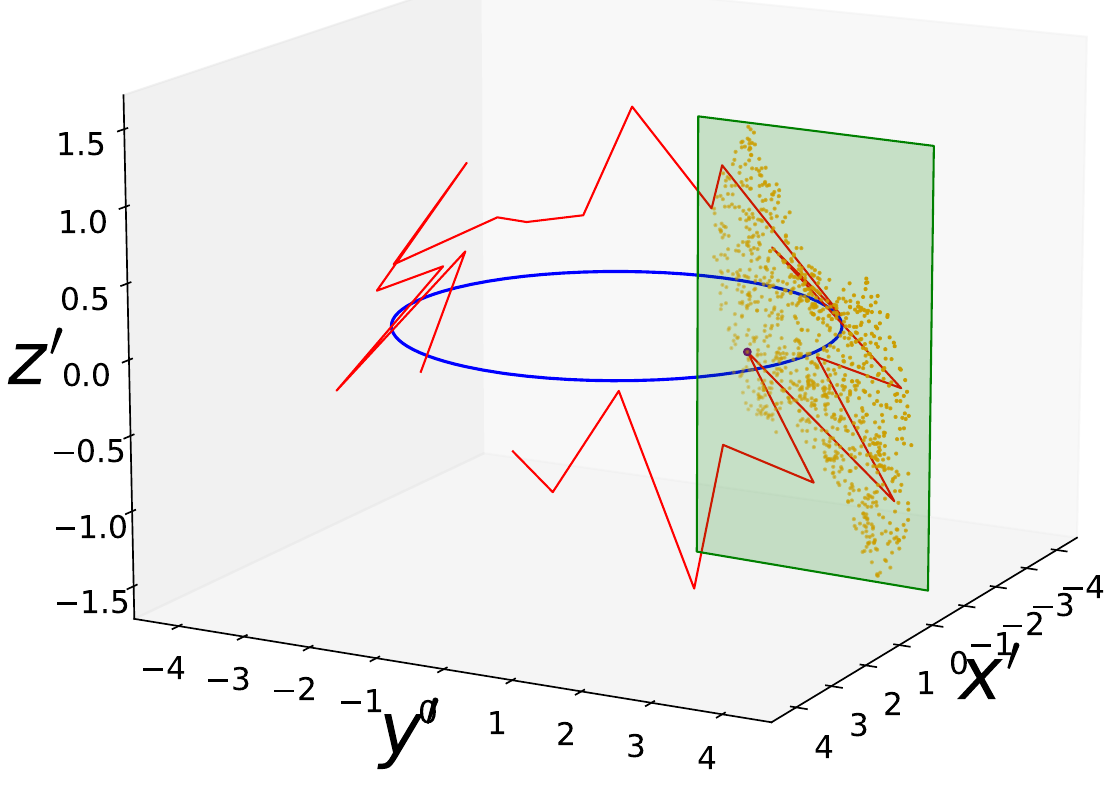}  
        \caption{}
        \label{fig:unscaled_traj}
    \end{subfigure}
    \hspace{30pt} 
    \begin{subfigure}[t]{0.45\textwidth}
        \centering
        \includegraphics[width=\textwidth]{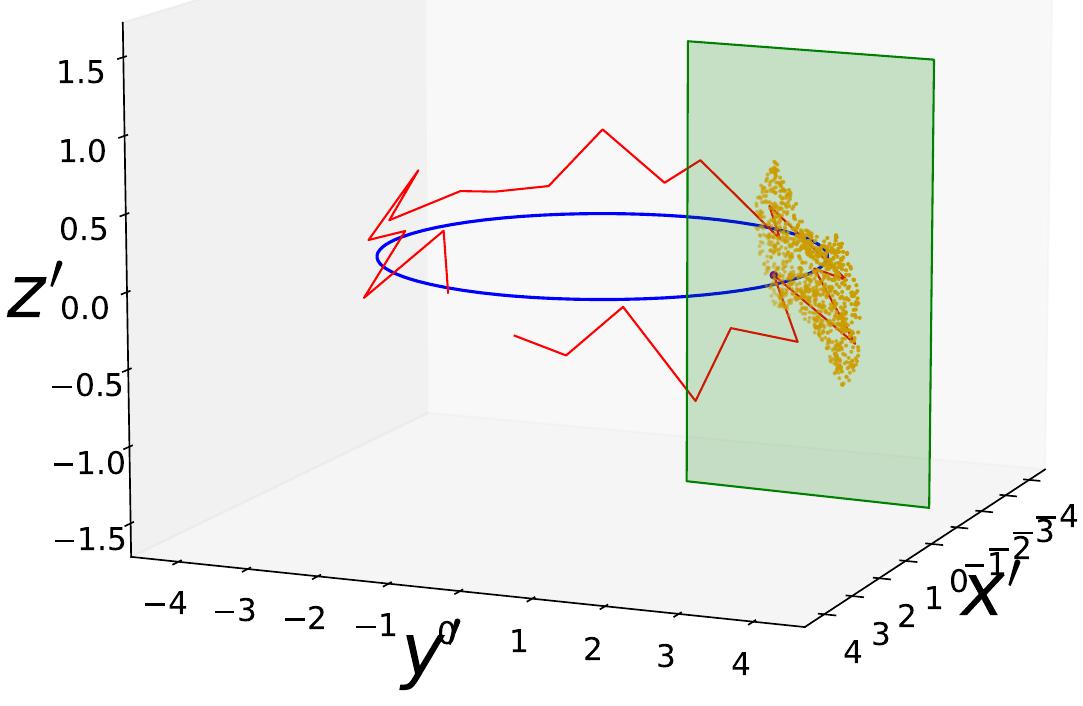}  
        \caption{}
        \label{fig:sc_traj}
    \end{subfigure}
    
    \vskip\baselineskip
    \begin{subfigure}[t]{0.45\textwidth}
        \centering
        \includegraphics[width=\textwidth]{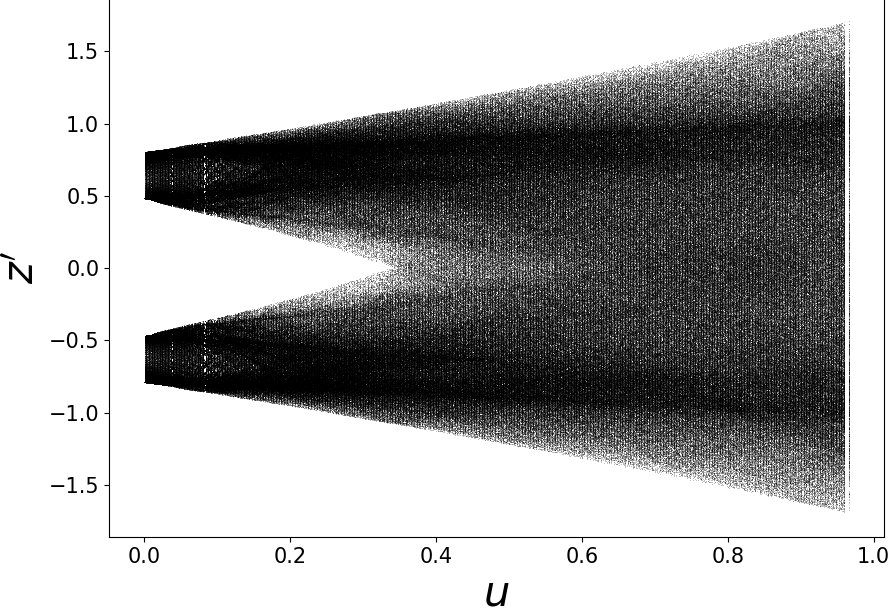}  
        \caption{}
        \label{fig:usca_bif}
    \end{subfigure}
    \hspace{30pt} 
    \begin{subfigure}[t]{0.45\textwidth}
        \centering
        \includegraphics[width=\textwidth]{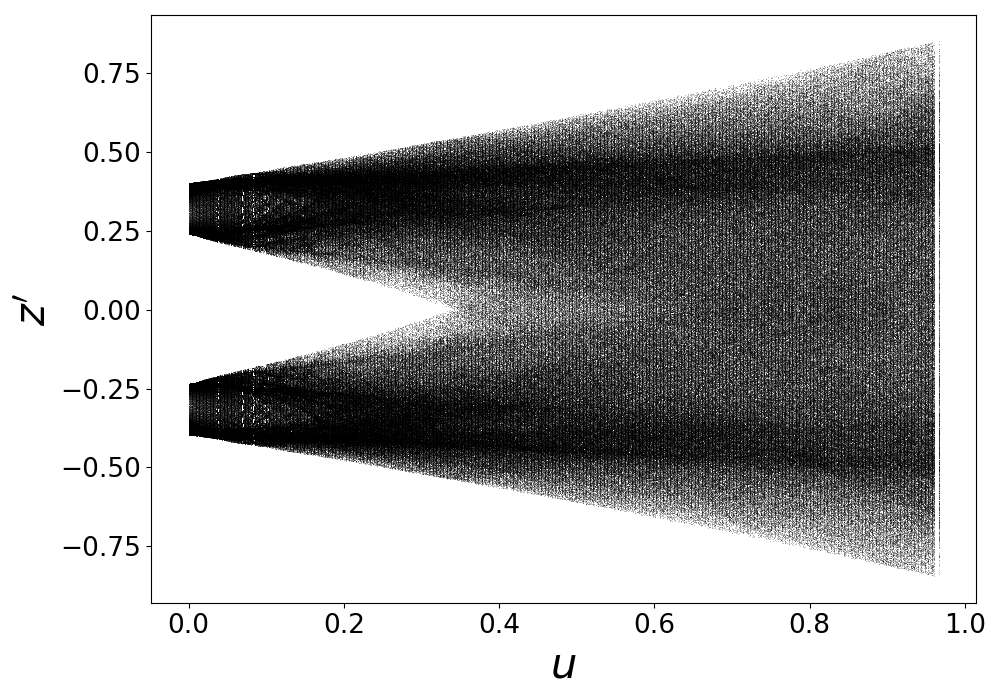}  
        \caption{}
        \label{fig:sc_bif}
    \end{subfigure}
    
    \caption{Illustration of scaling property of NDMH map. \textbf{a)} Plot of the unscaled hyperchaotic trajectory where \( S = 1 \)  \textbf{b)} The plot of scaled Trajectory where \( S = 0.5 \), which exhibits lesser proximity around the contour. \textbf{c)} Bifurcation plot for \( S = 1 \), where \( z' \) ranges from -1.65 to 1.65. \textbf{d)} Bifurcation plot for \( S = 0.5 \), where the range of variation of \( z' \) is from -0.82 to 0.82.}
    \label{fig:scal_plots}
\end{figure*}

In that case, this paper suggests that after scaling, a hyperchaotic map still retains both its Lyapunov exponents positive because of the affine nature of this transformation \cite{39}. Here, visual validation is provided through a closed circular contour of radius 3 and a step size of 25 for better clarity. In Fig.\ref{fig:unscaled_traj}, the seventh plane is plotted along the contour and mapped the 2d NDMH map on it for observing scaling properties. In this Fig.\ref{fig:unscaled_traj}, the sequence of hyperchaotic points is adapted by plotting the nth point on the nth plane to generate a trajectory shown as red lines. In Fig.\ref{fig:usca_bif}, it illustrates the bifurcation plot of the 2D NDMH map determined with respect to the global z' coordinate and iterated for each value in the range of the control parameter u from 0 to 1. The global reference frame is anchored at the origin of the circular contour, with no orientation of the contour relative to the z' global axis. Consequently, in the bifurcation plot Fig.\ref{fig:usca_bif}, the trajectory variation in the z' direction is shown, which indicates proximity around the contour spans from -1.65 to 1.65  with no scaling applied or S = 1 in Eq.\ref{eq:scaled_matrix}.

In Fig.\ref{fig:sc_traj}, the scaled trajectory with \(S = 0.5\) is plotted along the contour and shows that the map in this plane is significantly less covered than for the unscaled version. Hence in Fig.\ref{fig:sc_bif}, the range of variation of \( z' \) in the bifurcation diagram reduced to \(-0.82\) to \(0.82\), in contrast to the earlier unscaled trajectory, which ranged from \(-1.65\) to \(1.65\). The proximity distance was then reduced from 3.30 to 1.65 units.
This adaptation of scaling factor S ensures that the quadrotor would not leave the specified proximity around the contour, and also varying S can adjust the proximity for necessities like obstacle avoidance. This approach is particularly useful in scenarios where constraints in proximity need consideration, which happens to be common in practical applications.

\begin{figure}[htbp]
    \centering
    \begin{minipage}{0.41\textwidth}
        \centering
        \includegraphics[width=\textwidth]{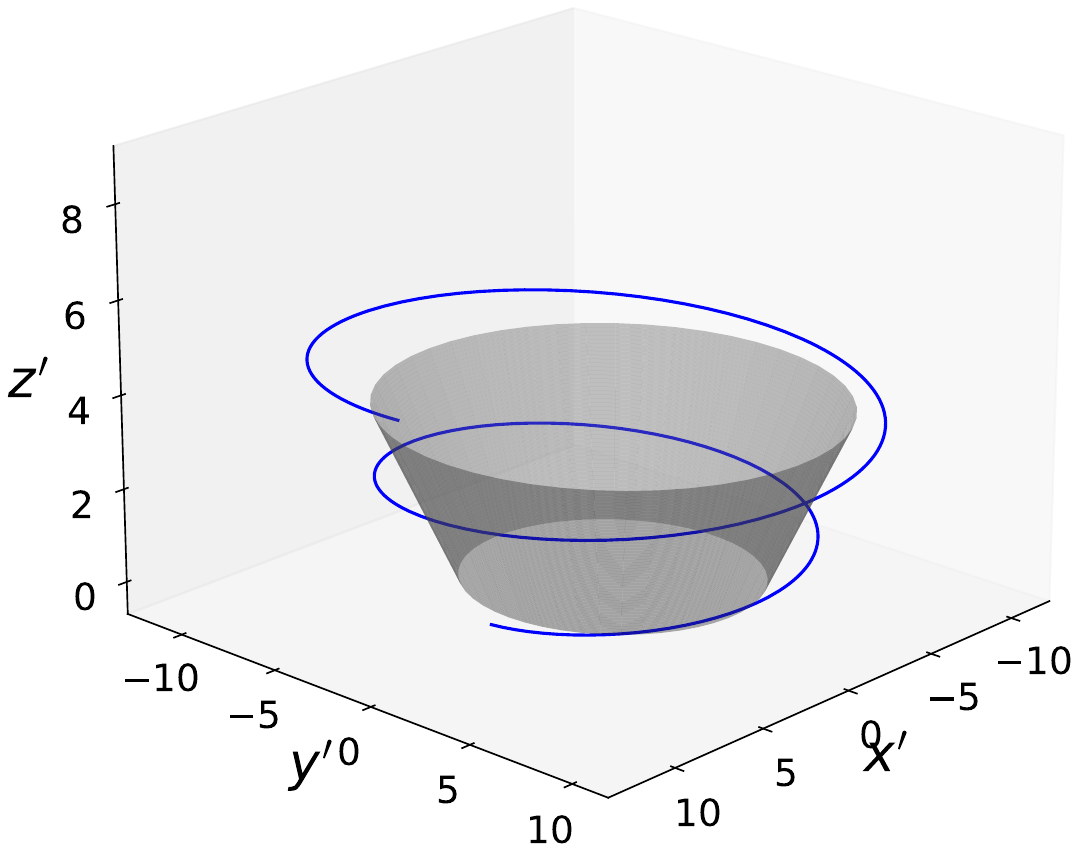}
        \subcaption{}
        \label{fig:stad_contour}
    \end{minipage}
    \hspace{0.04\textwidth}
    \begin{minipage}{0.41\textwidth}
        \centering
        \includegraphics[width=\textwidth]{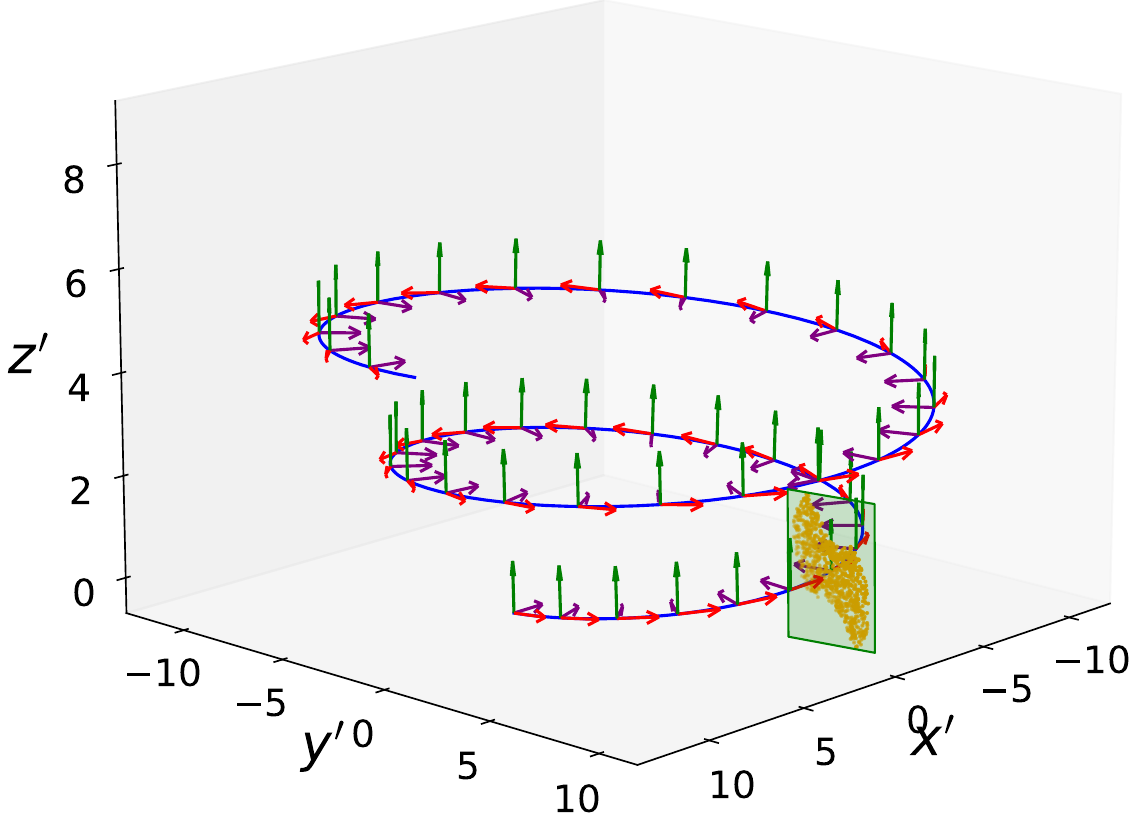}
        \subcaption{}
        \label{fig:mov_cords}
    \end{minipage}

    \begin{minipage}{0.447\textwidth}
        \centering
        \includegraphics[width=\textwidth]{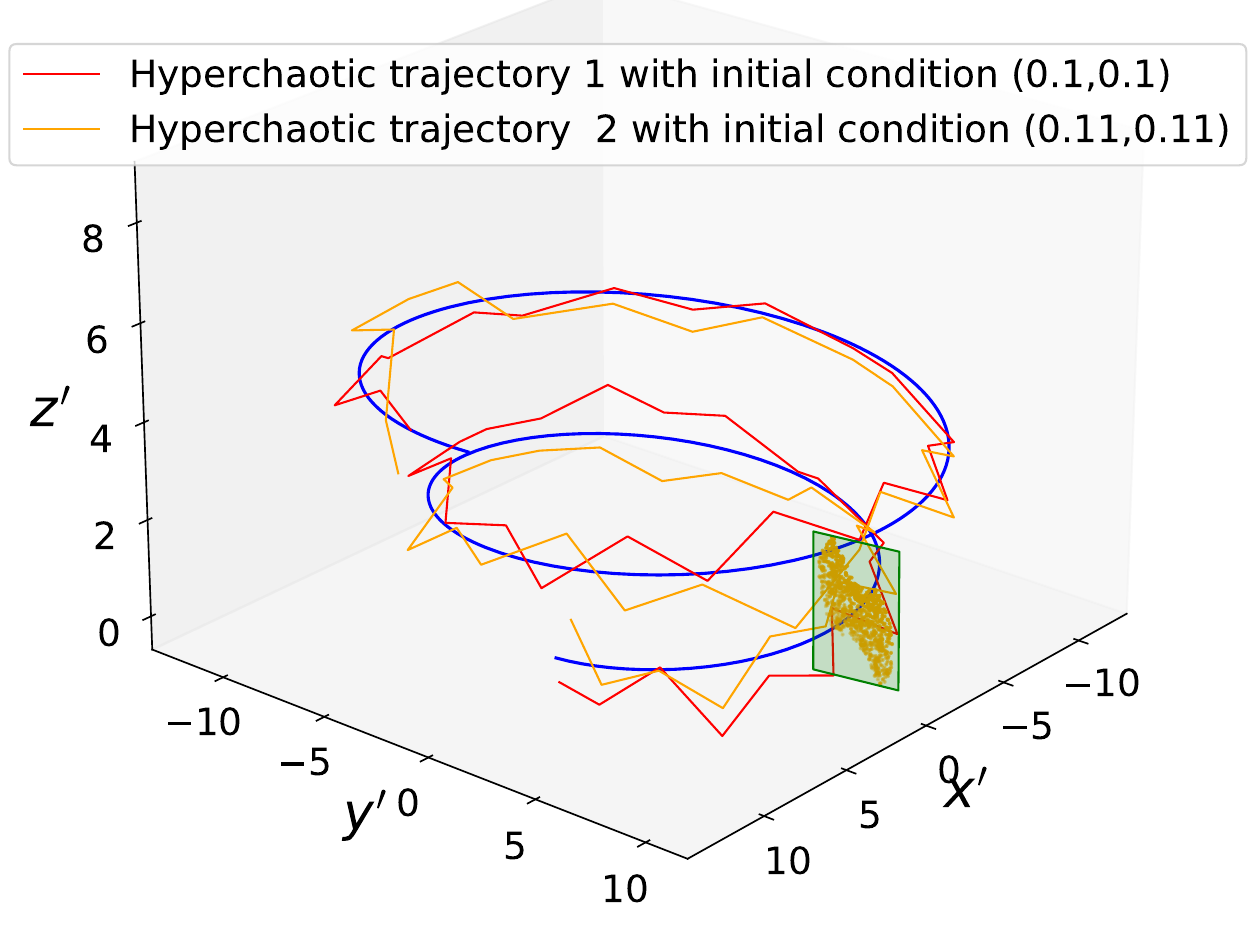}
        \subcaption{}
        \label{fig:two_traj}
    \end{minipage}

    \begin{minipage}{0.43\textwidth}
        \centering
        \includegraphics[width=\textwidth]{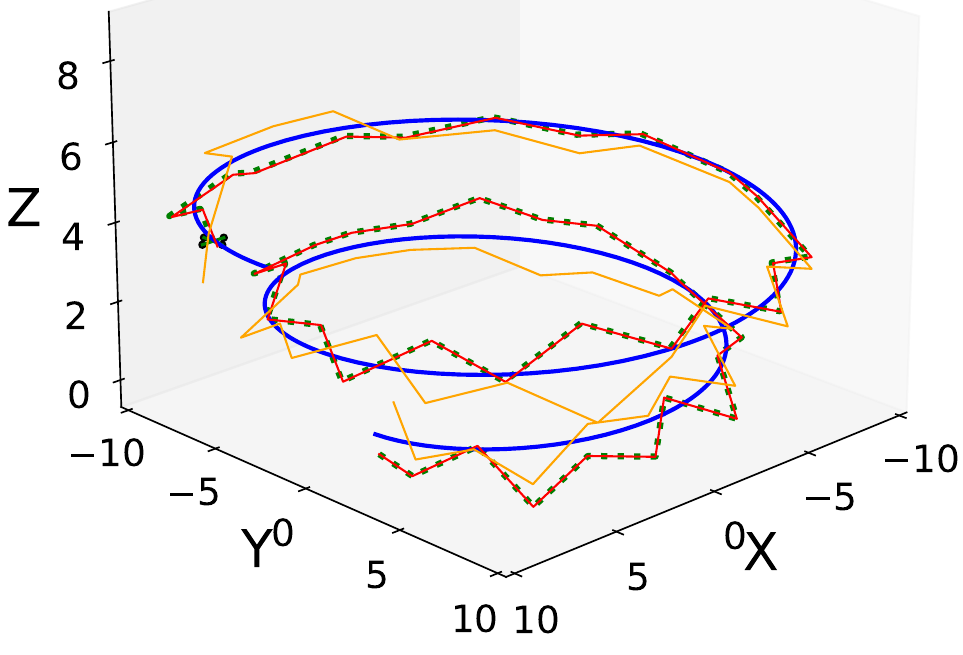}
        \subcaption{}
        \label{fig:QRT1}
    \end{minipage}
    \hspace{0.04\textwidth}
    \begin{minipage}{0.43\textwidth}
        \centering
        \includegraphics[width=\textwidth]{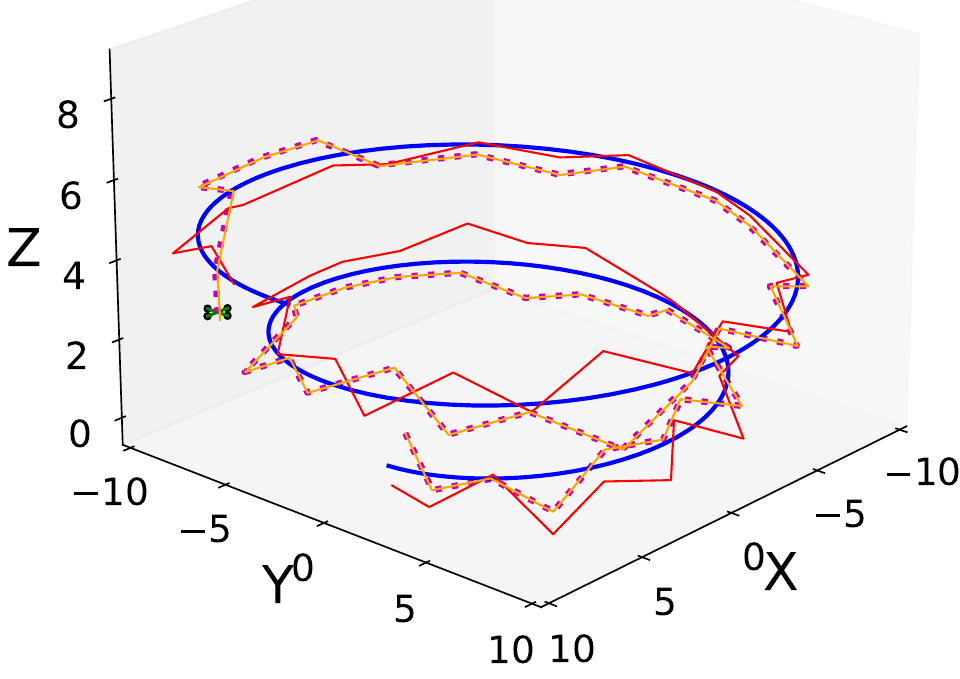}
        \subcaption{}
        \label{fig:QRT2}
    \end{minipage}
    
    \vskip\baselineskip
    \begin{subfigure}[t]{0.3\textwidth}
        \centering
        \includegraphics[width=\textwidth]{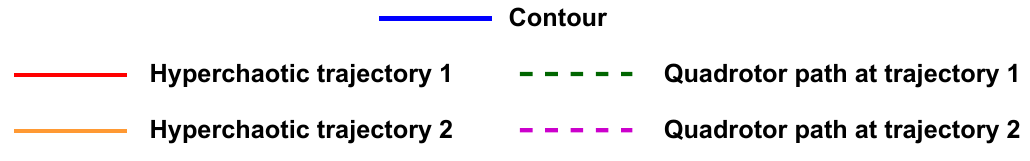}  
        
    \end{subfigure}

    \caption{Two Hyperchaotic trajectories and quadrotor tracking.\textbf{a)} Showing a helical contour with incremental radius. \textbf{b)} a 2D NDMH map is plotted along the contour with moving coordinates.\textbf{c)}  Two hyperchaotic trajectories with initial conditions (0.1, 0.1) and (0.11, 0.11) and shown in red and orange color.\textbf{d)} Plot of quadrotor tracing the first trajectory. \textbf{e)}  Plot of quadrotor tracing the second trajectory.}
    \label{fig:10}
\end{figure}

The objective of this paper is to derive multiple quadrotor hyperchaotic trajectories that are mutually independent, which enables multiple quadrotors for boundary surveillance around vertical structures. This is due to the independence and divergence properties of the chosen NDMH hyperchaotic system. Surveillance around a stadium-like structure with two turns of helical contour progressing vertically along the Z' axis was considered and is shown in Fig.\ref{fig:stad_contour}. In Fig.\ref{fig:mov_cords}, the moving coordinates are plotted with simulation parameter steps, which is set to 40, and this value was chosen for better visual observation. A single plane is also plotted along the contour to illustrate that the trajectory is part of a 2D NDMH map.
In Fig.\ref{fig:two_traj}, two hyperchaotic trajectories with slightly different initial conditions  (0.1,0.1) and (0.11,0.11) are plotted. Trajectory 1 and Trajectory 2 are represented by the red and orange lines, respectively. Both hyperchaotic trajectories were started after 49,000 iterations to discard transient behaviour and achieve higher divergence. Further trajectories included may overcrowd Fig.\ref{fig:two_traj}, hence making it difficult to understand.

In Fig.\ref{fig:QRT1}, the green dotted line shows the trajectory traced by the simulated quadrotor model. An efficient control system has been utilized to minimize the noise as well as accumulated errors for precise adherence to the hyperchaotic trajectory. The quadrotor tracks the desired trajectory without significant deviations due to the fact that the control parameter $T_P$ is set to a trajectory duration of 60. Larger values of $T_P$
 provide greater stability and slower joining paths. Therefore, a slower profile was chosen for quadrotor
tracing. Also, Fig.\ref{fig:QRT2} shows the quadrotor following the second trajectory, with the magenta dotted line indicating its path traced.

\section{Real-world implications}
\label{real_world_implications}
Sudden changes in the quadrotor's trajectory at waypoints can result in high variation of linear and angular acceleration in very short periods, which introduces extra noise in acceleration and gyroscope sensors\cite{60}. This is because accelerometers are naturally very susceptible to mechanical oscillations due to internal moving parts. It is possible to mitigate this issue by maintaining a stable distance between planes, ensuring smooth and stable quadrotor motion. This will, in turn, help in efficient tracing of the hyperchaotic trajectory. Some other methods, like complex electronic circuits and filtering algorithms, are also available, which can be applied for noise reduction.

\section{Conclusion}
\label{conclusion}
This paper presents the hyperchaotic path planning for a quadrotor to do boundary surveillance missions under adversarial conditions. From  2D contours, upgrading to helical 3D contours can substantially improve the effectiveness and unpredictability of boundary surveillance and enhance coverage while flying adaptively around vertical structures like buildings. Based on our algorithm, the rotation and sliding of 2D hyperchaotic maps create 3D trajectories along the helical contour. The proposed method provides a significant advance from previous approaches of using chaotic systems in 2D robot path planning to 3D applications by introducing a hardware-realizable hyperchaotic NDMH map. And parametrically evaluated the high divergence of the  NDMH system with others. In real-world applications, considering obstacle avoidance and constraints around boundaries are crucial. Since the hyperchaotic system supports scaling affine transformation, it can also serve that purpose by trajectory adaptation. The results of this article show that the generated hyperchaotic trajectories are mutually independent. So a quadrotor can trace these hyperchaotic trajectories with high precision if an effective control system is employed. More importantly, the methodology proposed to generate numerous hyperchaotic trajectories allows multiple quadrotors to trace them simultaneously. So operating thousands of quadrotors within a defined boundary is almost impossible to achieve manually, but our approach makes it feasible autonomously.  This approach makes hyperchaotic trajectories to enable a swarm of quadrotors to function concurrently for efficient coverage of complicated surveillance tasks while maintaining autonomy and scalability. Future work will focus on refining the scalability of the hyperchaotic trajectory algorithm for larger quadrotor swarms and integrating advanced mechanisms for obstacle avoidance that could ensure higher real-time adaptiveness and robustness in dynamic environments.
\bibliographystyle{unsrt} 
\bibliography{Arxiv}






\end{document}